\documentclass[]{aastex}

\usepackage{emulateapj5}
\usepackage{onecolfloat}
\usepackage{graphicx} 
\usepackage{fancyheadings} 
\usepackage{ulem}
\usepackage{rotating}
\usepackage{lscape}

\newcommand{\cm}[1]{\, {\rm cm^{#1}}}
\newcommand{\mkms}{{\rm \; km\;s^{-1}}}

\newcommand{\lya}{Ly$\alpha$}

\newcommand{\N}[1]{{N({\rm #1})}}
\newcommand{\sci}[1]{{\rm \; \times \; 10^{#1}}}
\newcommand{\mnhi}{N_{\rm HI}}

\newcommand{\afeg}{[$\alpha$/Fe]}
\newcommand{\nhii}{$\N{H^+}$}
\newcommand{\nhi}{$N_{\rm HI}$}
\newcommand{\nhtwo}{$N_{\rm H_2}$}
\def\fnhi{$f(\mnhi)$}
\def\mfnhi{f(\mnhi)}
\def\ltk{\left [ \,}
\def\ltp{\left ( \,}

\def\rtk{\, \right  ] }
\def\rtp{\, \right  ) }

\begin{document}
\twocolumn[%
\submitted{Submitted to ApJ; March 2 2007}

\title{Probing the ISM Near Star Forming Regions with GRB 
Afterglow Spectroscopy: Gas, Metals, and Dust}

\author{Jason X. Prochaska\altaffilmark{1},
        Hsiao-Wen Chen\altaffilmark{2}, 
	Miroslava Dessauges-Zavadsky\altaffilmark{3},
        Joshua S. Bloom\altaffilmark{4}
}

\begin{abstract}
We study the chemical abundances of the interstellar medium 
surrounding high $z$ gamma-ray bursts (GRBs) through analysis of the damped 
\lya\ systems (DLAs) identified in afterglow spectra.
These GRB-DLAs are characterized by large \ion{H}{1} column densities
\nhi\  and metallicities $\lbrack$M/H$\lbrack$ spanning
1/100 to nearly solar, with median $\lbrack$M/H$\lbrack$~$>-1$\,dex.
The majority of GRB-DLAs have $\lbrack$M/H$\lbrack$ values exceeding the cosmic mean
metallicity of atomic gas at $z>2$, i.e.\ if 
anything, the GRB-DLAs are biased to larger 
metallicity. 
We also observe 
(i) large $\lbrack$Zn/Fe$\lbrack$ values ($>+0.6$\,dex) and sub-solar Ti/Fe ratios
which imply substantial differential depletion,
(ii) large $\alpha/$Fe ratios suggesting nucleosynthetic enrichment by 
massive stars, 
and 
(iii) low C$^0$/C$^+$ ratios ($<10^{-4}$).
Quantitatively,
the observed depletion levels and C$^0$/C$^+$ ratios 
of the gas are not characteristic of cold, dense \ion{H}{1} clouds 
in the Galactic ISM.  We 
argue that the GRB-DLA represents the ISM near the GRB
but not gas directly local to the GRB (e.g.\ its molecular cloud
or circumstellar material).
We compare these observations with DLAs 
intervening background quasars (QSO-DLAs). 
The GRB-DLAs exhibit larger \nhi\ values, higher $\alpha$/Fe and Zn/Fe
ratios,  and have higher metallicity than the QSO-DLAs.
Although these differences are statistically significant, the offsets
are relatively modest (\nhi\ excepted).  We argue that the differences
primarily result from galactocentric radius-dependent
differences in the ISM: GRB-DLAs preferentially
probe denser, more depleted, higher metallicity gas located in the
inner few kpc whereas QSO-DLAs are more likely to intersect the 
less dense, less enriched, outer regions of the galaxy.
Finally, we investigate whether dust obscuration may exclude GRB-DLA
sightlines from QSO-DLA samples; we find that the majority of GRB-DLAs
would be recovered which implies little observational bias against
large \nhi\ systems.

\altaffiltext{1}{Department of Astronomy and Astrophysics, 
UCO/Lick Observatory;
University of California, 1156 High Street, 
Santa Cruz, CA 95064; xavier@ucolick.org}
\altaffiltext{2}{Department of Astronomy; University of Chicago;
5640 S. Ellis Ave., Chicago, IL 60637; hchen@oddjob.uchicago.edu}
\altaffiltext{3}{Observatoire de Gen\`eve, 51 Ch. des Maillettes, 
1290 Sauverny, Switzerland}
\altaffiltext{4}{Department of Astronomy, 601 Campbell Hall, 
University of California, Berkeley, CA 94720-3411}


\end{abstract}
]

\section{Introduction}

For the past decade, high resolution spectroscopy of distant
quasars (QSOs) have enabled detailed studies of the chemical abundances
of high $z$ galaxies \citep[e.g.][]{lu96,pw99,mirka01,molaro01,pro01,ledoux03}.
These properties of the interstellar medium (ISM) have been surveyed extensively 
via the damped \lya\ systems (DLAs), galaxies intersected by sightlines to
distant quasars \citep[for a review, see][]{wgp05}.
These QSO-DLA sightlines sample galaxies according to  
gas cross-section, i.e. the outer regions
of a galaxy's ISM are more frequently probed because differential
cross-section is proportional to radius.  
If star formation (SF) occurs in compact regions in
high $z$ galaxies (i.e.\ molecular clouds), then
QSO-DLAs would only rarely probe these environments \citep{zp06}.
This conclusion holds true 
independent of any additional biases related to dust obscuration by the
SF regions themselves \citep[e.g.][]{fall93}.  
QSO-DLA surveys are an inefficient approach to directly studying
the gas in high $z$ SF regions.

Like quasars, gamma-ray bursts (GRBs)
provide a bright -- albeit transient -- light beam from the
distant universe.  Imprinted on the roughly
power-law spectrum of the afterglow are the signatures of the
intergalactic medium \citep[e.g.][]{fdl+05,cpb+05,ppc+06}
and transitions from the ISM surrounding the
GRB event \citep[e.g.][]{sff03,vel+04,pcb+07}.  
And, analogous to absorption line studies of quasars, these
observations reveal the \ion{H}{1} column density
\citep{vel+04,jfl+06}, metallicity \citep{fsl+06,pro06},
chemical abundances, differential depletion \citep{sf04}, and
kinematics \citep[Paper\,II;][]{pcw+07} of the gas along the sightline.

The progenitors of long-duration GRB are believed to be massive
stars residing in active star forming regions \citep{woo93,wb06}.   
At low redshift,
this association is well established: supernova have
been identified at the same position as GRB events 
\citep[e.g.][but see also \citet{fwt+06}]{hsm+03,smg+03,mha+06}. 
Furthermore, GRBs have been found recently to be located 
within Wolf-Rayet galaxies \citep{hfs+06}.
The connection between GRB and star-forming regions
is inferred at high redshift. GRBs are found
exclusively in galaxies that are blue and show
nebular emission lines indicative of
ongoing star formation \citep{ldm+03}.
Furthermore, high precision astrometry of long-duration GRB afterglows 
reveal these events occur exclusively within 10\,kpc
of the flux-weighted centroid of their host galaxy \citep{bkd02,fls+06}.
Therefore, the sightlines to GRBs
should probe the ISM within or at least near star-forming regions. 
In this respect, studies of the ISM along sightlines to GRBs
complement those toward quasars.

The first observations of GRB-DLAs indicated large \ion{H}{1}
column density \citep{cgh+03,vel+04} and large metal-line equivalent
widths and column densities \citep{mdk+97,sff03}.  These data
suggested modestly enriched gas ($\gtrsim 1/10$ solar) and
substantial depletion levels but the analysis is limited
by line-saturation \citep[i.e.\ low-resolution spectroscopy of lines with
large optical depth;][]{pro06}.  With the launch of the Swift satellite
\citep{geh00},
GRBs are detected at a rate of $\approx 2$ per week with rapid, precise localizations
enabling echelle observations of a modest sample of GRB-DLAs 
\citep{cpb+05,pcb06,vls+07,pcb+07,pwf+07}.
These observations provide precise column density measurements of many
metal-line transitions and permit the 
analysis of the chemical abundances in 
the gas surrounding GRB. 

In this paper, we describe the chemical abundances 
of a modest sample of damped \lya\ systems associated
with the ISM surrounding gamma-ray bursts.  
We focus on echelle observations acquired in the past two
years, but also include pre-Swift observations.
Our principal goal is to describe the 
physical conditions of the interstellar medium within and/or 
near star formation regions in the young universe.
To frame the discussion, we will compare the GRB-DLA observations
against similar sets of observations for damped \lya\ systems
along quasar sightlines (QSO-DLAs).

There are at least three reasons why one may expect the ISM
characteristics of the GRB-DLAs to differ with those of QSO-DLAs.
First, the two samples may arise from overlapping yet
non-identical populations of high $z$ galaxies.
By definition, the QSO-DLAs correspond to galaxies with large \ion{H}{1}
surface densities.
These are selected independently of any emission or stellar property
\citep{wolfe86}
and cosmological simualtions suggest
a subset may not even be located within
dark matter halos \citep{rnp+06}.   Long-duration
GRB, in contrast, are known to occur only within star-forming galaxies
and are expected to roughly trace the on-going 
star formation rate \citep[SFR;e.g.][]{tot99,rtb02,gp07}.  
Of course, star-forming galaxies 
presumably also have
large \ion{H}{1} column densities\footnote{Similarly, we may expect that
all galaxies with large \ion{H}{1} column density will exhibit
star formation.  Indeed, the presence of heavy elements in all QSO-DLAs
indicates previous star formation \citep{pgw+03} and
\ion{C}{2}$^*$ fine-structure absorption implies ongoing
star formation \citep{wpg03}.},
as evidenced by the DLA profiles in GRB afterglows.

Second, even if QSO-DLAs and GRB-DLAs are drawn from the exact
same parent population of host galaxies,  the QSO-DLAs are 
selected according to \ion{H}{1} covering fraction on the sky while
GRB are expected to track current SFR.  One may expect
QSO-DLA to preferentially arise in galaxies with extended
\ion{H}{1} disks (e.g.\ low-surface brightness) whereas GRB
preferentialy occur within high-surface brightness galaxies
\citep{mmw98}.
Finally, and perhaps the most important, QSO-DLAs will 
preferentially probe the outer regions of galaxies
whereas GRB-DLAs are generally located
within the inner few kpc of their host galaxies \citep{bkd02}.
This could imply higher \nhi\ values, 
metallicity, depletion levels, molecular fractions, and
differential rotation along the GRB-DLAs sightlines.
Therefore, a secondary theme of this paper is to 
test the hypothesis that
GRB-DLAs and QSO-DLAs are drawn from the same parent population of high
$z$ galaxies and that any differences in ISM properties
are explained by their average impact parameters through the galaxies.

This paper is organized as follows.
Section~\ref{sec:exp} presents the experimental design,
and observational samples. 
In $\S$~\ref{sec:gas}, we characterize the
\ion{H}{1} column densities, metallicities, relative abundances,
and atomic carbon abundance of GRB-DLAs.
We discuss the implications of these results in 
$\S$~\ref{sec:discuss} and conclude with a list of 
future directions.

\section{Experimental Design and Observational Samples}
\label{sec:exp}

\subsection{Sightline Geometries}
\label{sec:geom}

Before discussing the observational data, it is important to comment
further on a few fundamental differences between studying DLAs along
GRB sightlines versus background QSOs.
The most obvious differences are that the GRB sightline
originates within the galaxy giving rise to the damped \lya\ system
and, almost certainly, from within a young star-forming region.
This implies several important consequences for the DLAs
probed by GRB afterglows.  

\begin{table*}[ht]\footnotesize
\begin{center}
\caption{{\sc GRB-DLA SAMPLE\label{tab:grbobs}}}
\begin{tabular}{lcccccccc}
\tableline
\tableline
GRB & RA &  DEC & $z_{GRB}$ & Instrument & 
$R$ & \ion{Mg}{1}$^a$ & Exc. \ion{Fe}{2}$^b$ & Ref \\
\tableline
GRB990123&15:25:30.34&+44:45:59.1&1.600&Keck/LRIS&1,000&Y&N&1\\
GRB000926&17:04:09.00&+51:47:10.0&2.038&Keck/ESI&5,000&Y&N&2\\
GRB010222&14:52:12.55&+43:01:06.2&1.477&Keck/ESI&5,000&Y&N&3\\
GRB011211&11:15:17.98&-21:56:56.2&2.142&VLT/FORS2&1,000&?&N&1,4\\
GRB020813&19:46:41.87&-19:36:04.8&1.255&Keck/LRIS&1,000&Y&Y&5\\
GRB030226&11:33:04.93&+25:53:55.3&1.987&Keck/ESI&5,000&?&N&6\\
GRB030323&11:06:09.40&-21:46:13.2&3.372&VLT/FORS2&1,000&Y&N&7\\
GRB050401&16:31:28.82&+02:11:14.8&2.899&VLT/FORS2&1,000&?&N&8\\
GRB050505&09:27:03.20&+30:16:21.5&4.275&Keck/LRIS&1,000&?&N&9\\
GRB050730&14:08:17.14&-03:46:17.8&3.969&Magellan/MIKE&30,000&?&Y&10\\
GRB050820&22:29:38.11&+19:33:37.1&2.615&Keck/HIRES&30,000&N&N&11\\
GRB050904&00:54:50.79&+14:05:09.4&6.296&Subaru/FOCAS&1,000&?&?&12\\
GRB050922C&19:55:54.48&-08:45:27.5&2.199&VLT/UVES&30,000&W&Y&13\\
GRB051111&00:08:17.14&-00:46:17.8&1.549&Keck/HIRES&30,000&Y&Y&14,11\\
GRB060206&13:31:43.42&+35:03:03.6&4.048&WHT/ISIS&4,000&?&?&15\\
GRB060418&15:45:42.40&-03:38:22.80&1.490&Magellan/MIKE&30,000&Y&Y&11\\
\tableline
\end{tabular}
\end{center}
\tablenotetext{a}{Strong MgI absorption  (W=weak).  GRB-DLA with strong 
\ion{Mg}{1} absorption are likely to have the majority of their gas at
distances greater than 50pc from the afterglow \citep{pcb06}}.
\tablenotetext{b}{Positive detection of absorption from excited levels
of Fe$^+$.  This measurement is difficult in low-resolution data and
a non-detection should be interpreted as ambigous.  The absence of 
absorption in higher resolution data, however, indicates the Fe$^+$ gas 
(observed via resonance lines) is at large distance from the afterglow.}
\tablerefs{
1: \cite{sff03};
2: \cite{cgh+03};
3: \cite{mhk+02};
4: \cite{vsf+06};
5: \cite{bsc+03};
6: \cite{sbp+06};
7: \cite{vel+04};
8: \cite{wfl+06};
9: \cite{bpck+05};
10: \cite{cpb+05};
11: \cite{pcb+07};
12: \cite{kka+06};
13: \cite{pwf+07};
14: \cite{pcb06};
15: \cite{fsl+06}}
 
\end{table*}

First, because the phenomena originates from within the DLAs themselves, 
a GRB sightline intersects less gas column than a 
background QSO sightline would at the same position on the sky.
Therefore, one underestimates the average \ion{H}{1} column density
and may underestimate the total
velocity field that occurs along the sightline.
Second,  it is possible that gas local to the progenitor, 
i.e.\ circumstellar material and/or molecular cloud gas,
contributes significantly to the GRB-DLAs (but see below).  One may
expect especially large dust-to-gas ratios, high molecular fractions,
and/or peculiar chemical abundance ratios.
Third, the GRB afterglow radiation can significantly affect the DLAs observed
along its sightline whereas QSO-DLAs are selected to have large
distance from their background QSO.  The UV flux from a GRB afterglow
that is typical of spectroscopic observations
(i.e.\ peak magnitude of $R \approx 15$ at early times) is sufficient
to ionize an \ion{H}{1} column of approximately $10^{21} \cm{-2}$
at 5\,pc from the afterglow in only a few minutes time
\citep[e.g.][]{draine02,pl02,pcb06}.  Similarly,
theoretical treatments predict the X-ray and UV components of
the afterglow can destroy dust
and molecular hydrogen out to 10 to 100pc \citep{wd00,fkr01,draine02}.
If neutral gas and dust existed at distances of $\sim 10$pc
prior to the burst, it is likely removed by the afterglow before
spectroscopic observations are initiated.  In this respect, the 
contribution of gas local to the GRB is somewhat minimized 
by the GRB event itself.
By the same token, gas local to the star forming regions may have been 
attenuated by the afterglow radiation field
and our observations
may not precisely reflect the conditions of this gas in the
absence of a GRB. 

On the other hand, aside from the UV pumping of fine-structure lines
\citep{pcb06,dcp+06,vls+07}, there is no compelling case of a
GRB afterglow having attenuated its surrounding medium. 
There have been no substantiated
reports of line variability in any resonance line including, 
and most importantly, \ion{Mg}{1} transitions.
The majority of GRB-DLAs exhibit strong
\ion{Mg}{1} absorption coincident in velocity with the resonance
and excited transitions of low-ion species \citep{pcb06,pcb+07}.  
This atom has an ionization potential IP=7.7\,eV and would be ionized
by the afterglow if it occurs within $\approx 100$\,pc of the event.
\cite{pcb06} have argued, therefore, that the majority of neutral
gas observed in GRB-DLAs is at a distance exceeding 100\,pc from the
progenitor.  Meanwhile, the detection of UV pumped excited states
of Si$^+$, O$^0$, and Fe$^+$ require that the gas is located within
$\approx 1$\,kpc of the afterglow.  Indeed, \cite{vls+07} have
analyzed variations in the level populations of the excited levels
of Fe$^+$ and Ni$^+$ and constrain the distance of the GRB-DLAs along
the GRB~060418 sightline to be at $1\pm 0.2$\,kpc from the afterglow.

\begin{figure}[ht]
\plotone{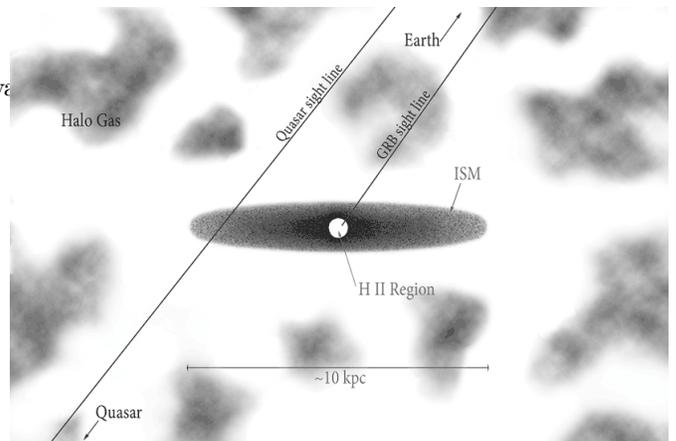}
\caption{This cartoon illustrates the likely differences
between QSO-DLA and GRB-DLA sightlines.  The former have randomly
intersected a foreground galaxy. These QSO-DLA sightlines correspond
to a cross-section selected sample and should preferentially
intersect the outer regions of the ISM in high $z$ galaxies.
In contrast, the GRB-DLAs are constrained to originate from within
the ISM of their host galaxies, presumably the \ion{H}{2} region
produced by massive stars in a star-forming region.  These
GRB-DLA sightlines are expected (and observed) to originate within
the inner few kpc of the ISM.
}
\label{fig:cartoon}
\end{figure}

\begin{table*}[ht]\footnotesize
\begin{center}
\caption{\sc {QSO-DLA SAMPLES\label{tab:qsosub}}}
\begin{tabular}{llccc}
\tableline
\tableline
Sample & Description & N$_{\rm DLA}$ & Instrument(s) & Ref \\
\tableline
HR-A&  High-resolution full sample (all) & 153 & HIRES, ESI, UVES & 1,2,3,4,5 \\
HR-S& High-resolution `statistical' sample & 112 & HIRES, ESI, UVES & 5 \\
HR-E & High precision echelle & 71 & HIRES, UVES & 1,2 \\
SDSS  & True (\nhi) statistical sample & 475 & SDSS  & 6,7 \\
\tableline
\end{tabular}
\end{center}
\tablerefs{
1: \cite{pwh+07};
2: \cite{dz06};
3: \cite{ledoux06};
4: \cite{shf06};
5: \cite{pgw+03};
6: \cite{ph04};
7: \cite{phw05}}
\end{table*}

Table~\ref{tab:grbobs} indicates those GRB-DLAs that exhibit
strong \ion{Mg}{1} absorption.  We also identify the GRB-DLAs
which show absorption from excited states of Fe$^+$.  Those cases
without Fe$^+$ fine-structure absorption must also lie at distances
$\gtrsim 100$\,pc \citep{pcb06,cpb+07}.  
These results indicate that the majority
of gas in GRB-DLAs is not local to the star formation region
and should not be adversely affected by the GRB afterglow.
Our current expectation is that O and B stars from the 
star-forming region have ionized all gas within $\approx 100$\,pc
of the GRB producing large \ion{H}{2} regions.
Figure~\ref{fig:cartoon} presents a cartoon illustration of 
a GRB and a QSO sightline through a galaxy comprised of a neutral ISM
surrounded by predominantly ionized halo gas clouds.  The
GRB sightline originates near the center of the galaxy,
within an \ion{H}{2} region associated with ongoing star-formation.
In contrast, the QSO-DLA sightlines will preferentially penetrate 
the outer regions of the ISM.

Independent of the differences one may expect for QSO-DLAs
and GRB-DLAs,  there may be common trends
(e.g.\ metallicity vs.\ dust depletion) characteristic
of the ISM of all high $z$ galaxies.  By considering the QSO-DLAs
and GRB-DLAs together, we hope to reveal these global trends
in addition to highlighting differences in the samples.

\subsection{GRB-DLA and QSO-DLA Samples}
\label{sec:samples}

The QSO-DLA samples are drawn from two sources: (i) high-resolution
echelle and echellette
observations of quasars acquired with the HIRES \citep{vogt94}
and ESI \citep{sheinis00} spectrometers at the Keck Observatory
and the UVES \citep{uves} spectrometer at the VLT Observatory;
and 
(ii) the low-resolution QSO-DLA surveys of the Sloan Digital Sky Survey 
\citep[SDSS;][]{ph04,phw05}.
We will restrict the samples to $z_{DLA} > 1.6$ and DLAs which are
greater than $3000\mkms$ from their background quasar.  The high-resolution 
sample (HR-A) is summarized in these papers: \cite{shf06,ledoux06,dz06,pwh+07}.
\cite{pwh+07} have emphasized that the full dataset is a heterogeneous
sample of QSO-DLAs including a number of systems selected on
the basis of strong metal-lines \citep{shf06} or as promising candidates
for H$_2$ absorption \citep{ledoux03}.  As such, the 
\ion{H}{1} distribution \fnhi\ of the high-resolution sample
does not follow the statistical distribution derived from the SDSS.

For a number of the comparisons made in this paper, we will take
subsets of these samples.  In particular, we define a pseudo-statistical
sample of metallicity measurements (HR-S) which is
restricted to the compilation of \cite{pgw+03}.  
We also construct a sample of high quality echelle-only
observations (HR-E) for analysis related to relative abundances
(e.g.\ $\alpha$/Fe).  The echellette observations, in 
particular, have too poor data quality to generally achieve better than
0.1\,dex precision \citep{p03_esi} and are not included.
Although the DLAs in these high-resolution 
subsets do not follow \fnhi\ for a random sample,
they were selected only on the basis of 
a large \ion{H}{1} column density.
Table~\ref{tab:qsosub} summarizes the various QSO-DLA samples
considered in this paper.

Our selection criteria for the GRB-DLA sample are
(i) the presence of a damped \lya\ system ($\mnhi \ge 2\sci{20} \cm{-2}$)
or a low-ion column density that requires
\nhi\ exceed $2 \sci{20} \cm{-2}$ assuming solar metallicity\footnote{We note
that the few GRB sightlines with $\mnhi < 10^{20} \cm{-2}$ 
have very low column densities for low-ion transitions \citep{fdl+05,gcn3971}.};
(ii) spectra with sufficient signal-to-noise and resolution 
to study the chemical abundances of the gas.
The sample is also limited to GRB-DLAs where
we could access the data or where precise equivalent
width measurements were reported in the literature.
Table~\ref{tab:grbobs} lists the GRB-DLAs comprising
our sample and describes the spectral observations.

As with QSO-DLAs \citep{fall93}, it is likely 
that dust obscuration 
plays a role in defining the GRB-DLA sample.  There are 
several examples in the literature of highly reddened afterglows
which are likely extinguished by dust in the GRB 
host galaxy \citep[e.g.][]{lfr+06,pdl+06}. 
At present, no comprehensive study on the incidence
of optically `dark' bursts has been performed nor an evaluation
of the fraction of dark bursts which are cases of dust obscuration
(as opposed to high $z$ events).  We expect, however, that dust
obscuration is important to defining GRB-DLA samples.

\subsection{Absorption-Line Metallicities}


Metallicity is the mass density in metals relative to hydrogen and helium gas.  
It is likely that oxygen dominates this quantity in nearly
every astrophysical environment, with carbon and nitrogen 
secondary.  Although these three elements exhibit
low-ion transitions, the majority are either too strong, too
weak, and/or located within the \lya\ forest.
Therefore, the metal abundance of the interstellar medium is
frequently gauged by other elements.
Unfortunately, many of the transitions which are accessible to analysis
arise from elements that are refractory.    To avoid the complications
of depletion corrections (which can be an order of magnitude or more),
observers have focused on non-refractory or mildly refractory elements.
These include S, Si, and Zn. The latter is a trace element,
log(Zn/H)$_\odot$ = $-7.3$, with an 
uncertain nucleosynthetic origin \citep{hwf+96} and, therefore,
should be considered cautiously.

A point of great interest in GRB studies is the metallicity
of the progenitor especially in the context of the collapsar model \citep{w93}.  
In the collapsar paradigm, 
large angular momentum is required to power the GRB.  
Because high metallicity stars are expected to have significant
mass-loss by winds promoting the loss of angular momentum \citep{vd05},
\cite{ln06} and \cite{wh06} have argued that GRB progenitors
will have low metallicity, i.e.\ less than 1/10 solar abundance.
One may consider this to be a natural prediction
of the collapsar model.

There is at least circumstantial evidence in support of a
metallicity `bias' from studies of $z<1$ GRB host galaxies.
The metallicities of the \ion{H}{2} regions in a small sample of 
$z<0.5$ galaxies hosting GRBs are sub-solar 
\citep{pbc+04,sof+05} and several authors have
noted that the values are systematically lower than the
predicted distribution for galaxies 
drawn randomly according to current SFR \citep{sgb+06,kbg+07,mkk+07}.
Furthermore, the GRB host galaxies at $z \sim 1$
are generally sub-L$_*$ 
consistent with low metallicity \citep[][]{ldm+03}.
\cite{fls+06} compared a sample of 
$z\lesssim 1$ GRB host galaxies against the host 
population for core-collapse
supernova and demonstrated that GRB host galaxies have 
systematically lower luminosity.  The difference is approximately 
one magnitude which they suggested could be due to a metallicity
bias.  However, galaxies follow
a metallicity-luminosity relation that is roughly linear \citep{kk04} and
a one magnitude difference in luminosity implies only an approximately
0.3\,dex offset in metallicity.  Recently, \cite{wp06} performed a thorough
analysis showing the observations suggest a
metallicity `ceiling' for GRB progenitors but that this cutoff
occurs at no lower than 1/2 solar.
Therefore, while the offset in luminosities between the host
galaxies of GRB and core-collapse supernova may be explained
by a metallicity bias, this does not imply as severe an effect
as promoted by studies of mass-loss.

At high redshift, it is difficult to observe nebular lines to
measure metallicities and
our knowledge of galaxies is too poor to make robust statements
from luminosity distributions alone.  Instead, one can measure the metallicity 
of the interstellar medium from absorption-line spectroscopy 
by comparing the total hydrogen column density with the total
column density of metals.

\section{Gas Phase Abundances in GRB-DLAs}
\label{sec:gas}

In this section, we present column densities of hydrogen
and metals observed in the gas-phase in GRB-DLAs.
These observations give the surface density and metallicity of
the ISM, and the relative abundances reflect differential depletion
and the underlying nucleosynthetic patterns in the gas.
Finally, we comment on the abundance of atomic carbon.
In all of this analysis, we draw comparisons between the GRB-DLAs
and QSO-DLAs.

\begin{sidewaystable*}\footnotesize
\begin{center}
\caption{\sc {GRB-DLA ABUNDANCE SUMMARY\label{tab:abnd}}}
\begin{tabular}{lccccccccccccccccccccccc}
\tableline
\tableline
GRB & log \nhi & 
f$_\alpha^a$ & [$\alpha$/H] &
$\sigma(\alpha$)$^b$ &
f$_{Zn}^c$ & [Zn/H] &
$\sigma$(Zn)$^b$ &
f$_{M}^d$ & [M/H] & $\sigma$(M)$^b$ & f$_{Fe}^e$ & [Fe/H] &
$\sigma$(Fe)$^b$ & [Ti/H]$^f$ & f$_{N}^c$ & [N/H] & $\sigma$(N) & $\N{C^0}^g$ \\ 
\tableline
GRB990123&$22^{a}$& 0&$$&& 2&$-0.97$&L.L.& 12&$-0.97$&L.L.& 2&$-2.00$&L.L.&& 0&$$&&\\
GRB000926&$21.30^{+0.25}_{-0.25}$& 2&$-1.58$&L.L.& 1&$-0.17$&0.15&  2&$-0.17$&0.29& 5&$-1.49$&0.08&-0.43& 0&$$&& 13.9\\
GRB010222&$22^{a}$& 2&$-1.61$&L.L.& 2&$-1.30$&L.L.& 12&$-1.30$&L.L.& 1&$-2.01$&0.08&-2.35& 0&$$&&\\
GRB011211&$20.40^{+0.20}_{-0.20}$& 2&$-1.36$&L.L.& 0&$$&& 11&$-1.36$&L.L.& 2&$-1.80$&L.L.&& 0&$$&&\\
GRB020813&$22^{a}$& 2&$-1.31$&L.L.& 2&$-1.17$&L.L.& 12&$-1.17$&L.L.& 2&$-2.10$&L.L.&-1.94& 0&$$&&\\
GRB030226&$20.50^{+0.30}_{-0.30}$& 2&$-1.31$&L.L.& 3&$-0.47$&U.L.& 11&$-1.31$&L.L.& 1&$-1.05$&0.18&& 0&$$&&\\
GRB030323&$21.90^{+0.07}_{-0.07}$& 2&$-1.56$&L.L.& 2&$-0.87$&L.L.& 12&$-0.87$&L.L.& 1&$-2.40$&0.43&& 0&$$&&\\
GRB050401&$22.60^{+0.30}_{-0.30}$& 2&$-2.16$&L.L.& 2&$-1.57$&L.L.& 12&$-1.57$&L.L.& 2&$-2.30$&L.L.&& 0&$$&&\\
GRB050505&$22.05^{+0.10}_{-0.10}$&-4&$-1.25$&L.L.& 0&$$&& 11&$-1.25$&L.L.& 2&$-1.35$&L.L.&& 0&$$&&\\
GRB050730&$22.15^{+0.10}_{-0.10}$& 4&$-2.26$&0.10& 0&$$&&  4&$-2.26$&0.14& 1&$-2.50$&0.12&& 1&$-3.16$&0.10& 13.2\\
GRB050820&$21.00^{+0.10}_{-0.10}$& 4&$-0.63$&0.04& 1&$-0.71$&0.02&  4&$-0.63$&0.11& 1&$-1.60$&0.09&-0.80& 2&$-1.35$&L.L.& 12.7\\
GRB050904&$21.30^{+0.20}_{-0.20}$&-4&$-1.10$&L.L.& 0&$$&& 11&$-1.10$&L.L.& 0&$$&&& 0&$$&&\\
GRB050922C&$21.60^{+0.10}_{-0.10}$& 4&$-2.03$&0.10& 3&$-2.02$&U.L.&  4&$-2.03$&0.14& 1&$-2.63$&0.01&-1.60& 3&$-4.09$&U.L.& 12.9\\
GRB051111&$22^{a}$& 2&$-1.42$&L.L.& 2&$-0.96$&L.L.& 12&$-0.96$&L.L.& 1&$-2.14$&0.01&-2.23& 0&$$&& 13.0\\
GRB060206&$20.85^{+0.10}_{-0.10}$& 4&$-0.85$&0.15& 0&$$&&  4&$-0.85$&0.18& 0&$$&&& 0&$$&&\\
GRB060418&$22^{a}$& 2&$-1.67$&L.L.& 1&$-1.65$&0.04&  2&$-1.65$&1.00& 1&$-2.24$&0.03&-2.23& 0&$$&& 12.9\\
\tableline
\end{tabular}
\end{center}
\tablenotetext{a}{0=No measurement; 1=Si measurement; 2=Si lower limit;
3=Si upper limit; 4=[S/H] ; 5=[O/H] ; 13=S+Si limits; -4=S lower limit}
\tablenotetext{b}{U.L. indicates upper limit, L.L. indicates lower limit.}
\tablenotetext{c}{0=No measurement; 1=Measurement; 2=Lower limit;
3=Upper limit}
\tablenotetext{d}{0=No measurement; 1=Si measurement; 2=Zn measurement;
3=Combination of limits; 4=S measurement; 11=Lower limit from [$\alpha$/H]; 12=Lower limit from Zn}
\tablenotetext{e}{0=No measurement; 1=Fe measurement; 2=Fe lower limit;
3=Fe upper limit; 4=[Ni/H]-0.1dex;
5=[Cr/H] - 0.2dex}
\tablenotetext{f}{Upper limit on the Ti abundance.}
\tablenotetext{g}{Upper limit on the column density of atomic carbon.}
\tablecomments{With the exception of [M/H], the errors reported in the [X/H] values only reflect uncertainty in X.}
 
\end{sidewaystable*}

\subsection{Hydrogen Column Densities}
\label{sec:hgas}

In principle, the absorption-line
spectra of rest-frame UV transitions ($\lambda < 3000$\AA)
will give measurements of the atomic hydrogen and molecular
hydrogen column densities, \nhi\ and \nhtwo. 
For a damped \lya\ system, the former is best determined
through a Voigt profile fit to the \lya~$\lambda 1215$
transition.  At $\mnhi > 10^{20} \cm{-2}$, the damping wings
of the Lorentzian line-profile are well-resolved with
a moderate resolution spectrum (FWHM$<5$\AA).  Therefore,
aside from the redshift, the \nhi\ value is the most readily 
measured physical characteristic of a damped \lya\ system.

\begin{figure}[ht]
\begin{center}
\includegraphics[height=3.5in,angle=90]{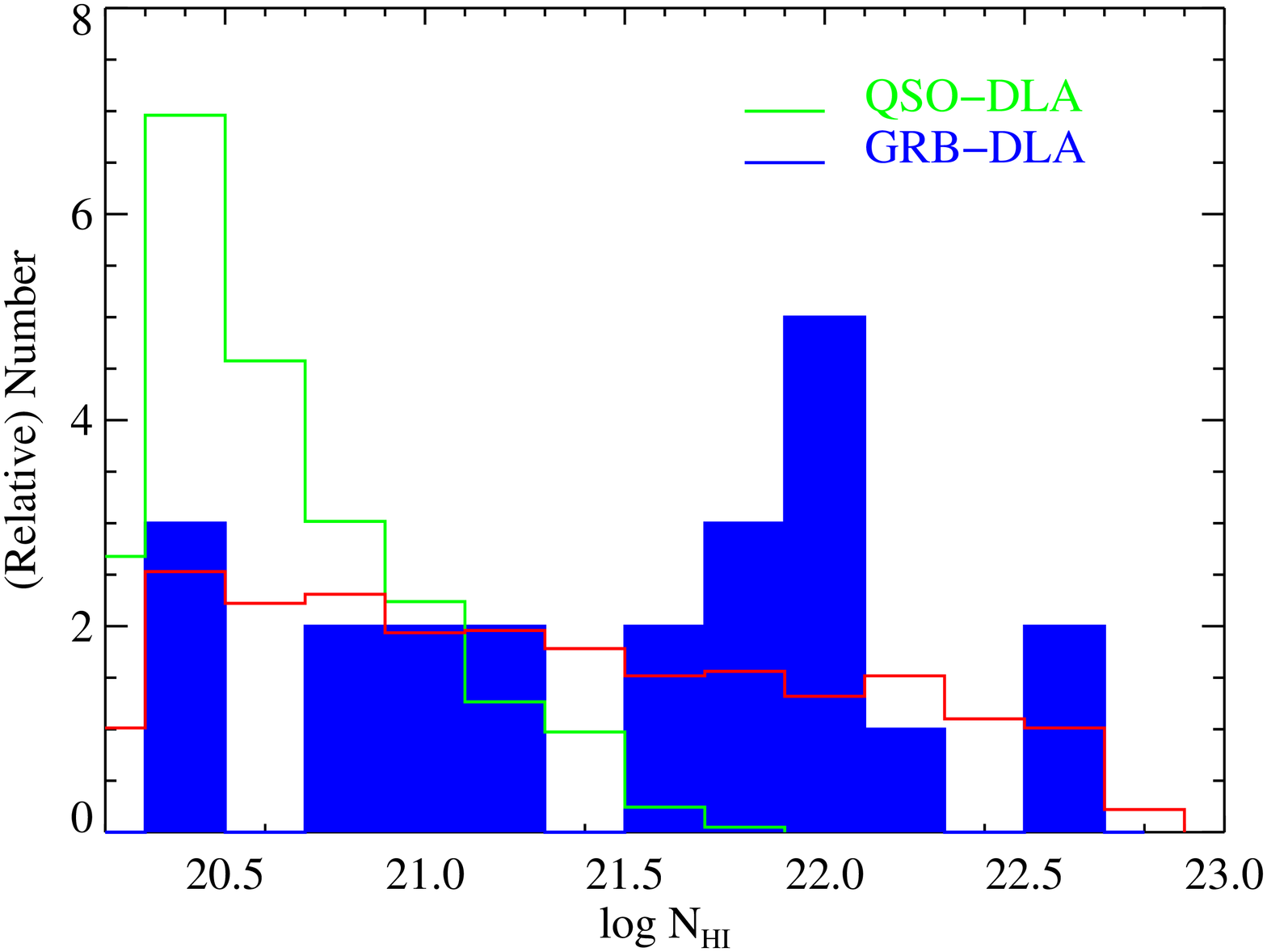}
\end{center}
\caption{Blue (dark, solid) curve traces the histogram
of \nhi\ values for the GRB-DLAs.   In comparison, we show
a histogram of \nhi\ values for QSO-DLAs drawn randomly toward
background quasars \citep{phw05} and normalized for presentation
to have the same number of systems as the GRB-DLA sample.
The GRB-DLAs have median value $\log \mnhi = 21.7$, which exceeds
all but a few QSO-DLAs observed to date.
The dotted red line traces the predicted distribution of 
\nhi\ values assuming a sample of sightlines originating inside
an \ion{H}{2} region located at the
center of an \ion{H}{1} exponential disk.  The model shown has central column 
density $\log N_0 = 22$, scale-height to scale-radius $h/R = 0.1$,
and an \ion{H}{2} region with radius $r_{HII} = 4h$.
}
\label{fig:nhihist}
\end{figure}

Figure~\ref{fig:nhihist} presents the sample of \nhi\
measurements for our GRB-DLA sample 
supplemented by the compilation of \cite{jlf+06}
and compared against the distribution of \nhi\ values for
QSO-DLAs measured from the Sloan Digital Sky Survey \citep{phw05}.
As previously reported \citep{vel+04,jlf+06}, the GRB-DLA
distribution is skewed to significantly higher \nhi\ values
than QSO-DLAs.  The difference, at least qualitatively,
supports the hypothesis that QSO-DLAs probe the
outer regions of galaxies whereas GRB-DLAs probe the 
inner, star-forming regions.  \cite{jlf+06} discussed that
the GRB-DLA distribution is roughly consistent with that predicted
for GRB embedded within molecular clouds \citep{rp02}.
The authors note, however, that GRB exhibit too many sightlines
with $\mnhi < 10^{22} \cm{-2}$ compared to the prediction
for molecular clouds.  They argue this may result
from photoionization and/or because GRB are preferentially 
located at the edge \citep{wnh+99} or even outside molecular
clouds \citep{hfs+06}.

As we discussed in $\S$~\ref{sec:geom}, however, the presence
of strong \ion{Mg}{1} absorption in nearly every GRB
sightline argues that the majority of GRB-DLA gas is 
located beyond $\approx 100$pc of the event.  This distance
is comparable to only the largest giant molecular cloud complexes
in the Local Group
\citep{bfk+07} and it is reasonable to assume that the GRB-DLA gas is generally
not associated with the molecular cloud hosting the GRB
(see $\S$~\ref{sec:grbdiscuss}).
Examining Figure~\ref{fig:nhihist} in this light, the most
salient question becomes: why do GRB-DLAs exhibit a preponderance
of $\mnhi > 10^{22} \cm{-2}$ measurements?  
Indeed, the gas mass required to average $\mnhi = 10^{22} \cm{-2}$
at 100\,pc along random sightlines
is $M_{HI} = 10^7 M_\odot$;  this exceeds the masses of even the
largest molecular clouds in the Milky Way \citep{sr89,blitz93}. 
It seems unlikely, therefore, that the observed \ion{H}{1}
gas corresponds to the circumstellar medium which hosted the GRB.
Instead, we contend the gas is associated with the nearby ISM of the galaxy.
This hypothesis, however, must account for the large \nhi\ values
while allowing for an evacuated volume (i.e.\ \ion{H}{2} region)
with radius 100\,pc or more.

To explore this point further, we considered an \ion{H}{1} disk
described by a double exponential\footnote{The following results are
not too sensitive to the functional form of the radial profile.},
\begin{equation}
n(Z,R) = \frac{N_0}{h} \exp \ltp \frac{-|Z|}{h} \rtp \exp 
\ltp \frac{-R}{R_d} \rtp \;\; ,
\end{equation}
with scale-height $h$, disk length $R_d$, and central \ion{H}{1} 
column density $N_0$.
We searched a wide parameter space of $h/R_d$, $N_0$, and radius of the 
\ion{H}{2} region $r_{HII}$.   Over-plotted on Figure~\ref{fig:nhihist}
is the prediction for random sightlines originating at the center
and mid-plane ($R=Z=0$) of an exponential disk 
with $\log N_0 = 22$, $h/R_d = 0.1$ and
$r_{HII} = 4h$.  This is a reasonably good description of the
observed distribution ($P_{KS} = 11\%$);  
an ambient ISM characterized by $\log N_0 = 22$ with a large
\ion{H}{2} region ($r_{HII} > 2h$) produces  
a reasonable model for the observed \nhi\ distribution.
It would be worthwhile to 
consider this model within the context of star forming galaxies
in cosmological simulations.

The association of GRB with 
star-forming regions raises the possibility that a significant
fraction of the gas in GRB-DLAs is molecular.
Conveniently, H$_2$ gives rise to several band heads
at $\lambda \approx 1000$\,\AA\ and one can directly measure
the H$_2$ column density.  For optical spectroscopy, this requires $z_{GRB} > 2$
and blue wavelength coverage.  It is also important to have 
relatively high resolution to
disentangle the absorption lines from the \lya\ forest.
\cite{tpd+07} have recently presented an analysis of five
GRB sightlines and set an upper limit to the molecular 
fraction $f({\rm H_2}) < 10^{-5}$ in four of the systems.  The only possible 
detection is along the sightline to GRB~060206 with 
$f({\rm H_2}) \approx 10^{-3.5}$ \citep{fsl+06}, and
even this value is likely to be an upper limit \citep{tpd+07}.
Therefore, we will assume that \nhtwo\ is a small fraction of the
total hydrogen column density. 
These results lend further support to the interpretation of 
GRB-DLAs as being dominated by ambient ISM instead of gas local to the
progenitor.

Lastly, there is the contribution to $\N{H}$ from H$^+$.  Out to distances of
several 10\,pc, the star formation region (or GRB
afterglow) will have produced an \ion{H}{2} region with a potentially
large \nhii\ value.  
That is, unless stellar winds from the GRB progenitor have nearly
evacuated the region altogether.
There is, of course, no
direct means of measuring the H$^+$ column density.   Instead, we restrict
our analysis of the metal column density to the atoms and ions
that are dominant in neutral regions.
These are termed ``low-ions'' -- Si$^+$, Fe$^+$, C$^+$, 
O$^0$.  We caution that low-ions are not entirely absent from \ion{H}{2} regions
\citep[e.g.][]{howk_ion99}, although the expected contribution would
be small compared to neutral gas for $\mnhi > 10^{21} \cm{-2}$.
A detailed treatment of the ionization state of GRB-DLAs will be
presented in a future paper. Here, we will 
assume that ionization corrections are small 
and comment on conclusions which are sensitive to this assumption.

\subsection{GRB-DLA Metallicities}
\label{sec:mh}

Ideally, the metal column density is derived from a single or set of 
unsaturated, resolved transitions of a low-ion.   
We will not attempt to measure the total column
density of each element (i.e.\ by summing all of the ionization
states of a given element) but only the state which is
dominant in neutral hydrogen regions (the low-ion).
This can generally be achieved
from spectra with high-resolution observations and modest signal-to-noise ratio.
To date, however, many GRB spectra have been acquired with low-resolution
spectrometers.  These observations may provide accurate measurements
of the equivalent widths (EWs) of strong lines, but precise column densities are 
difficult to derive due to modeling the line-profile \citep{jenkins86}.  
The high-resolution observations of GRB-DLAs, however, 
indicate that the metal-line profiles 
are comprised of multiple components (termed `clouds')
which exhibit a bimodal distribution of column densities \citep{pro06}.
In most cases, the total column density is dominated by a single or few
clouds whereas the equivalent width is the net sum of many weak clouds
(Paper~II).
When the cloud column densities exhibit a bimodal distribution,
the single-component COG analysis systematically 
underestimates the ionic column densities \citep{jenkins86,pro06}.
In the following, we will use low-resolution EW observations only
to set lower limits to column density measurements
and exclude these from
relative chemical abundances.


Table~\ref{tab:abnd} summarizes the abundance measurements for 
the GRB-DLAs.  We have corrected ions which exhibit fine-structure
splitting of the ground-state (e.g.\ Fe$^+$) by the observed
contribution of these levels.  In general, this implies an increment
of $<0.05$\,dex.
In five cases the \ion{H}{1} column density
and a non-refractory or mildly refractory metal abundance have been
precisely measured.  For the remainder of cases, we report a lower
limit to the metallicity because the metal abundance is a lower limit
and/or there is no spectral coverage of the \lya\ transition.  In the latter
cases, we have assumed $\mnhi = 10^{22} \cm{-2}$ which is approximately
the mean $\log \mnhi$ value of GRB-DLAs.

\begin{figure*}[ht]
\begin{center}
\includegraphics[height=6.5in,angle=90]{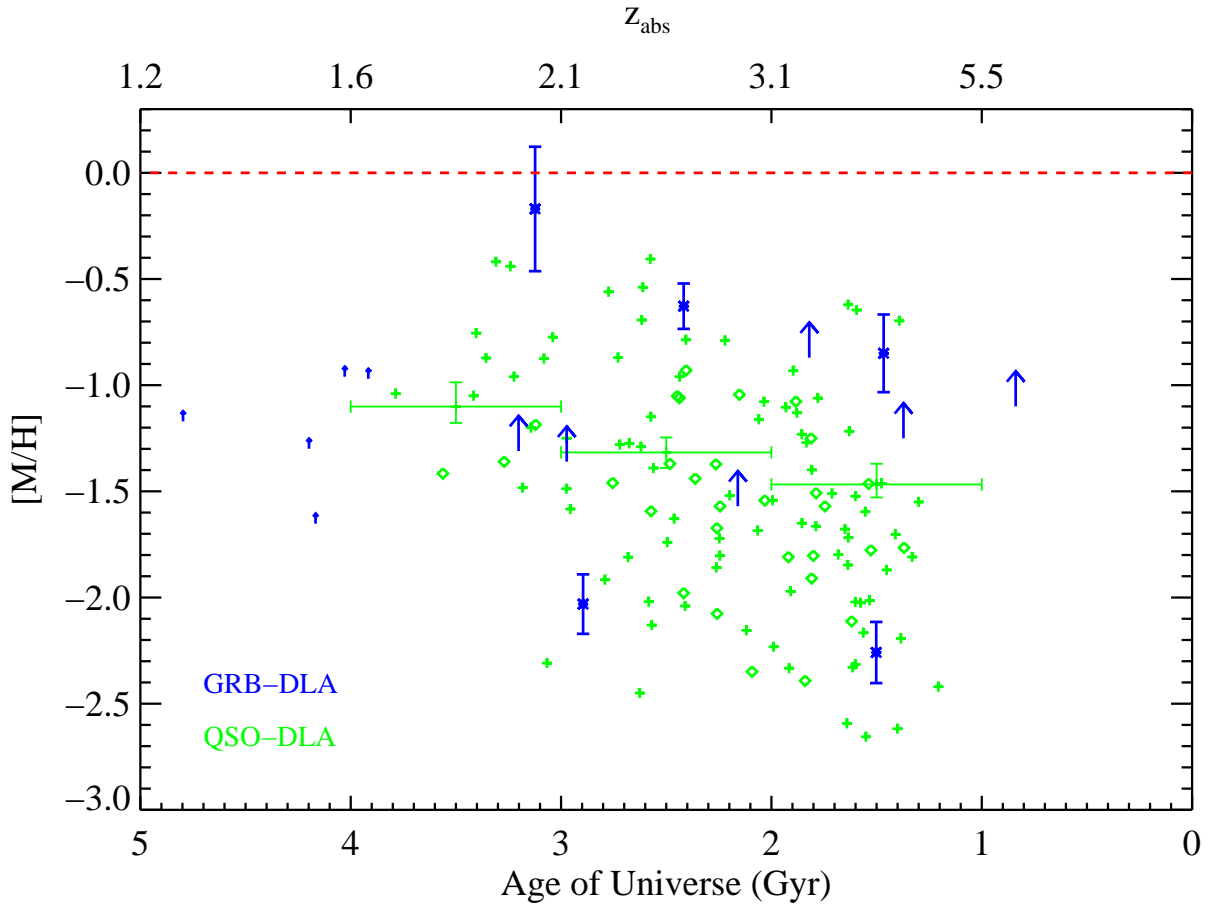}
\end{center}
\caption{Metallicity [M/H] measurements for the GRB-DLAs (dark blue) as
a function of the age of the universe corresponding to the observed
absorption redshift and assuming the current concordance cosmology \citep{wmap06}.
At $z<1.6$ where \lya\ is lost below the atmosphere, 
the small arrows indicate lower limits to the metallicity assuming 
$\mnhi = 10^{22} \cm{-2}$.  At $z>1.6$, the lower limits to [M/H]
for the GRB-DLAs are due to line-saturation.  The lighter points
show measurements for the QSO-DLAs assuming the HR-S sample \citep{pgw+03}.
The plus signs indicate $\mnhi < 10^{21} \cm{-2}$ and the diamonds correspond
to $\mnhi \ge 10^{21} \cm{-2}$.  We also present the cosmic mean
metallicity derived from the QSO-DLAs by taking the \ion{H}{1}-weighted
mean of the individual data points.  Comparing the two distributions,
we note that the majority of GRB-DLA values lie above the cosmic mean
and that a significant fraction have [M/H]~$>-1$.
}
\label{fig:mtl}
\end{figure*}

We have evaluated the metallicity of the gas by adopting
(in order of preference) the 
[S/H]\footnote{[X/Y] = $\log (X/Y) - \log (X/Y)_\odot$}
value, the [Si/H] value, the [Zn/H] value,
or the maximum of these if each has only a lower limit value.
We prefer S and Si to Zn because these elements are $\approx 1000\times$ more
abundant and because Zn has an uncertain nucleosynthetic origin \citep{hwf+96}.
Again, we avoid highly refractory elements like Fe and Ni (even though
they may dominate the opacity in stellar atmospheres and therefore
are particularly relevant within the collapsar model) because
these elements may be depleted from the gas-phase.

\begin{figure}[ht]
\begin{center}
\includegraphics[height=3.5in,angle=90]{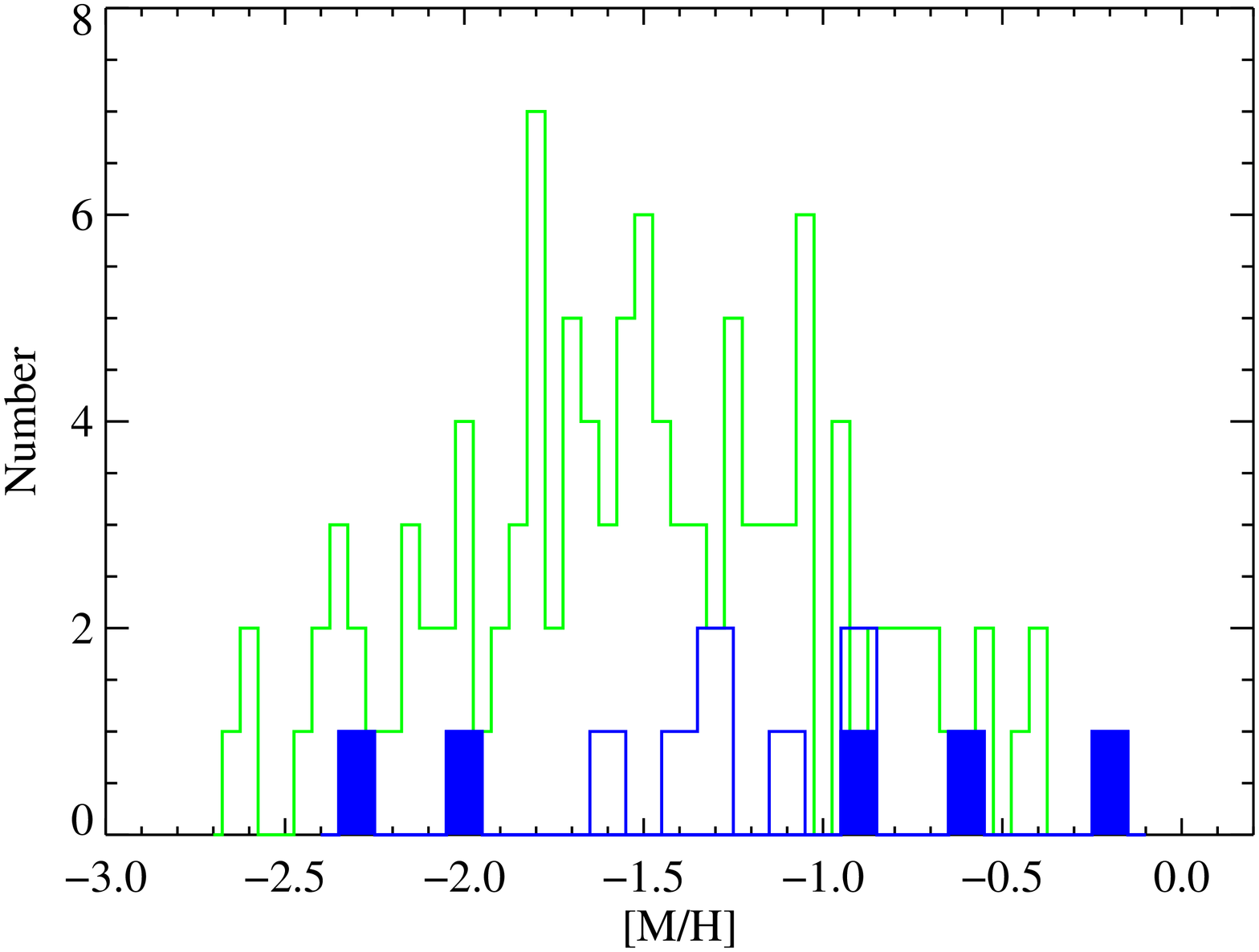}
\end{center}
\caption{Histogram of [M/H] measurements for the QSO-DLAs (light green)
and the GRB-DLAs (dark) with both samples restricted to $z>1.6$.  
For the GRB-DLAs, the open histogram traces
the lower limits to [M/H] because of line-saturation.  If these values are
evaluated as measurements, then the two distributions are consistent
with being drawn from the same parent population.  If we increment
the lower limits by +0.3\,dex, however, the null hypothesis is
ruled out at $>99\%$c.l. by a two-sided KS test.
Similarly, a two-sample survival analysis \citep{fn85} rules out the
null hypothesis at greater than 99$\%$c.l.
}
\label{fig:histmtl}
\end{figure}

Figure~\ref{fig:mtl} presents the metallicity measurements for the
GRB-DLAs as a function of the age of the universe (set by $z_{GRB}$)
assuming the concordance cosmology
\citep[$\Omega_\Lambda = 0.72, \Omega_m = 0.28, h=0.73$;][]{wmap06}.
Over-plotted on the figure are the metallicity measurements for the
statistical sample (HR-S) of QSO-DLAs. 
We restrict the discussion in this section to 
$t < 3.5$\,Gyr (i.e. $z>1.65$) where all of 
the GRB-DLAs have measured \nhi\ values. 
It is apparent that both the GRB-DLAs and
QSO-DLAs exhibit a large dispersion of values.  The peak-to-peak
range is $\approx 2$\,dex, much larger than observational 
uncertainty.  
Second, both samples exhibit a metallicity `floor' at approximately
1/1000 solar abundance.
This lower limit to the metallicities
is also not observational;  the sensitivity limit of the data 
is at least an order of magnitude lower.
It remains an open question whether this floor
is associated with early (PopIII) enrichment or rapid local enrichment
in all galaxies exhibiting DLAs \citep{wq00,pgw+03,qw03}.

Also over-plotted in Figure~\ref{fig:mtl} is the cosmic mean 
metallicity of atomic gas.  This quantity is derived from  
the \nhi-weighted mean of QSO-DLA metallicities \citep{lzt93,pgw+03}.
Six of the 10 GRB-DLAs with \nhi\ measurements exceed the cosmic
mean and several lower limits lie at the cosmic mean.
Therefore, many and likely most of the GRB-DLAs have metallicities
exceeding the cosmic mean in the ambient ISM of high $z$ galaxies.
On these grounds, at least, the GRB-DLAs at high $z$
do not appear to show a significant metallicity bias toward low values.  

Figure~\ref{fig:histmtl} presents a histogram of [M/H] values for the 
GRB-DLAs and QSO-DLAs (restricted to $z>1.6$).  The solid bars indicate
GRB-DLA values while the dark open bars show lower limits to [M/H] for
the GRB-DLAs.  We observe that
that the GRB-DLA values roughly overlap the QSO-DLA distribution.
A proper treatment of the observations
is to perform a two-sample survival analysis
\citep{fn85}.  Using the ASURV software package, we ran 
standard Gehan, logrank, and Peto-Peto tests using the Kaplan-Meier
estimator and found the null hypothesis is ruled out at 99$\%$c.l.
We conclude that the metallicities of GRB-DLAs are larger than those
for QSO-DLAs at $z>1.6$.

We note that many of the GRB-DLAs have a metallicity
greater than 1/10 solar.  At first glance, this appears to contradict
assertions that GRB progenitors have metallicity less than $0.1 Z_\odot$
\citep{ln06,wh06}.  We will demonstrate in the next
section, however, that the gas-phase abundances of S, Si, and Zn are
all enhanced relative to Fe.  The data allow, but do not require,
that the abundance of Fe is $\approx 0.5$\,dex lower than the 
[M/H] values.  It is possible, therefore, that at high $z$ the
GRB-DLA metallicities are not biased low and also that 
[Fe/H]~$<-1$ in most cases.
Finally, the gas that we observe is many 10\,pc's away from the progenitor
and need not accurately reflect its metallicity.
At present, however, we have no reason to suspect that the
ISM values would systematically overestimate the GRB progenitor
metallicity.
 
\begin{figure*}[ht]
\begin{center}
\includegraphics[height=6.5in,angle=90]{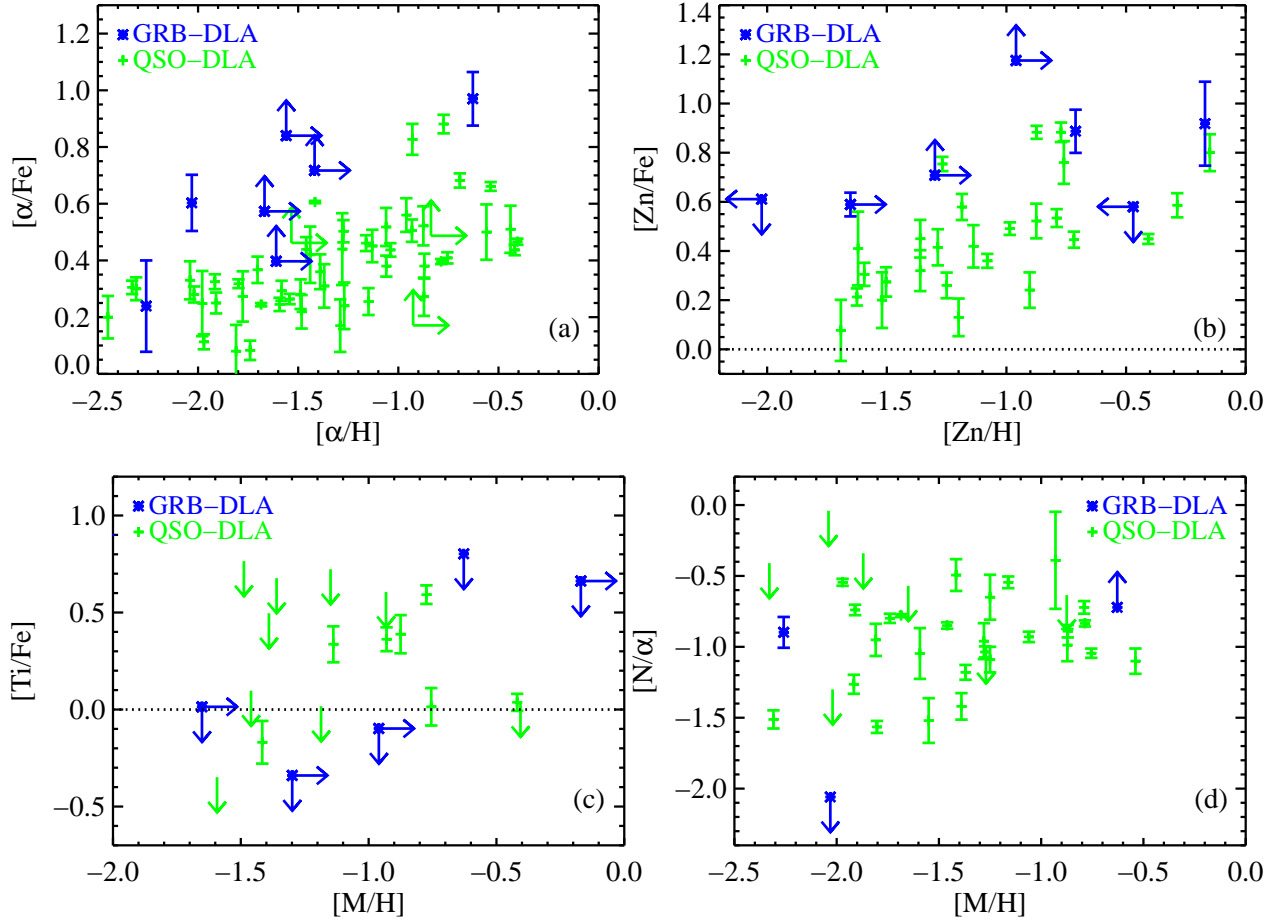}
\end{center}
\caption{Observed gas-phase relative abundances as a function of 
alpha abundance [$\alpha$/H], zinc abundance [Zn/H], or metallicity [M/H].
The four panels show the values for GRB-DLAs and QSO-DLAs (sample HR-E)
for (a) [$\alpha$/Fe] ratios, (b) [Zn/Fe] ratios, (c) [Ti/Fe] ratios,
and (d) [N/$\alpha$] ratios.  The GRB-DLAs are characterized by large
[$\alpha$/Fe] and [Zn/Fe] ratios indicating significant nucleosynthetic
enhancement by massive stars and/or differential depletion.  One
also notes several solar and sub-solar upper limits on [Ti/Fe] which
are strong evidence that refractory metals are depleted onto dust grains
in GRB-DLAs.
}
\label{fig:relabnd}
\end{figure*}

\subsection{Relative Abundances}

We now turn our attention to the relative abundances of
the GRB-DLAs.  These are measured by comparing the gas-phase 
column densities of pairs of low-ions (X$_i$, Y$_i$) under the assumption
that ionization corrections are small, i.e.
[X/Y] = $\log N_{X_i} - \log N_{Y_i} - \log(X/Y)_\odot$.
With high resolution observations and moderate S/N data,
one can frequently achieve 0.05\,dex precision or better for
$\log N_{X_i}$.
Because the observed ratios
represent gas-phase abundances, however, the values
reflect a combination of the underlying nucleosynthetic pattern
and the effects of differential depletion onto dust grains.
It is unfortunate that there is no element in the Fe peak 
which is non-refractory.  As such, observers frequently
employ Zn (a neighbor of the Fe peak) as a surrogate for Fe
because it is non-refractory and because it traces Fe in the Galaxy
at [Fe/H]~$>-2$ \citep{sgc91}.   We note, however, that Zn
has an uncertain nucleosynthetic origin and should not directly
trace Fe in galaxies whose star formation history
differs from the Milky Way \citep{fpg04}.
Therefore, we consider it to be a surrogate for Fe but with 
a large systematic uncertainty.

Throughout this section we restrict the QSO-DLA sample to the
high precision echelle measurements (sample HR-E; Table~\ref{tab:grbobs}).

\subsubsection{$\alpha$/Fe}

A key nucleosynthetic diagnostic of stars and galaxies is the 
$\alpha$/Fe ratio, where $\alpha$ refers to the sequence of He
fusion in massive stars, i.e. O, Mg, Si, S, Ar.
Standard theory of nucleosynthesis predicts these elements
are predominantly produced by massive stars \citep[e.g.][]{ww95}.
Therefore, a comparison of the $\alpha$ abundance with the Fe-peak
observations, whose
production is dominated by the Type\,Ia SN of less massive
stars, constrain the age and star formation history
of the galaxy when plotted against gas metallicity \citep{tinsley79}.
Unfortunately, O and Mg are difficult to measure via UV absorption
lines because the transitions are either too strong or too weak.
Instead, one typically estimates the $\alpha$ abundance
with Si or S.  In the QSO-DLAs, 
[S/Si]~$\approx 0$ and the elements are roughly 
interchangeable \citep{pw02}.  For the GRB-DLAs, we have also adopted these
two elements as the reference for $\alpha$.
The Fe-peak abundance, meanwhile, is determined from Fe, Ni, or Cr.  

Figure~\ref{fig:relabnd}a presents the observed $\alpha$/Fe
ratios for the GRB-DLAs against the QSO-DLAs as a function of 
$\alpha$ abundance, [$\alpha$/H].    
We reemphasize that these gas-phase abundances
have not been corrected for differential depletion.  
Consider, first, the values for the QSO-DLAs.  At low metallicity
($[\alpha/{\rm H}] < -1.5$), the QSO-DLAs follow a well-defined
`plateau' at \afeg$\approx 0.25$\,dex \citep{pw02}.  
This plateau matches the one observed for 
metal-poor Galactic stars \citep{mcw97} suggesting the $\alpha$/Fe
enhancement has a nucleosynthetic origin \citep{lu96,dz06}.
At higher metallicity,
the mean and dispersion of \afeg\ rises, in stark
contrast to the $\alpha$/Fe trend in the Milky Way and that
expected for other star formation histories 
\citep[e.g.][]{scg+02}.  The increase in  
$\alpha$/Fe with increasing metallicity is a clear signature for 
differential depletion \citep{pw02}, i.e.,
large \afeg\ ratios at high metallicity
result from the greater adsorption of Fe onto dust grains.

Turning to the GRB-DLAs, all of the \afeg\ measurements
are consistent with at least +0.5\,dex and the majority lie
at greater than $+0.6$\,dex.  
Two-sample survival analysis tests (e.g.\ Gehan) report a less
than $0.1\%$ probability that the \afeg\
values of the QSO-DLAs and GRB-DLAs are drawn from the same parent
population.  We conclude that the GRB-DLAs have systematically
higher \afeg\ ratios than the QSO-DLAs.
The question that follows is whether these higher values
indicate enhanced $\alpha$ abundances (nucleosynthesis)
and/or a higher depletion level for Fe (dust).

From the nucleosynthetic viewpoint, one expects enhanced 
$\alpha$/Fe in the gas near GRBs because 
(i) the progenitors
are massive stars and
(ii) the high specific star formation rates of their host galaxies  
imply ages that are young compared to the time-scales for
Type~Ia enrichment \citep{chg04}.
We contend that [$\alpha$/Fe]$>+0.3$\,dex
is strongly expected for GRB-DLA
gas from nucleosynthesis enrichment alone \citep[see also][]{cdp07}.   
Regarding dust, one may expect high depletion levels
for gas in or near star forming regions.
This could also explain the offset in $\alpha$/Fe between
the GRB-DLAs and QSO-DLAs, especially if GRB-DLAs have systematically
higher metallicity ($\S$~\ref{sec:mh}).

In the next subsection, we will consider dust in greater depth. 
Before proceeding, we wish to emphasize
the relatively low $\alpha$/Fe value for GRB~050730:
\afeg=$0.25 \pm 0.15$ at [$\alpha$/H]~$=-2.25$.
If we were to interpret this $\alpha$/Fe ratio in terms of depletion,
the intrinsic (nucleosynthetic) ratio would be approximately solar
or even sub-solar; this would require star-formation with low efficiency
over time-scales of a few 100\,Myr \citep{cmv03,dz07}.
This mode of star formation does not appear common for
high $z$ GRB host galaxies \citep{chg04}.
In this one case,  we argue that the gas
is essentially undepleted.  One may speculate further 
that the majority of QSO-DLAs at low metallicity are also
nearly undepleted and that their observed $\alpha$/Fe ratios
simply imply significant $\alpha$-enrichment.

\subsubsection{Differential Depletion}

It is routine in QSO-DLA studies to examine the Zn/Fe 
ratio to gauge the level of differential depletion
\citep[or Zn/Cr;][]{mr90,pettini94}.  As discussed above, this
assumes that the nucleosynthetic production of Zn tracks the Fe 
peak closely such
that enhancements in the gas-phase abundance of Zn/Fe 
is primarily due to Fe depletion.
While this assumption should be questioned at the level of a 
few tenths dex \citep[e.g.][]{pnc+00}, 
large Zn/Fe enhancements -- especially at high metallicity -- 
are unlikely to be explained by nucleosynthesis alone.

We present the Zn/Fe ratios of GRB-DLAs and QSO-DLAs
in Figure~\ref{fig:relabnd}b.  
Similar to the $\alpha$/Fe ratios,
the Zn/Fe values for the QSO-DLAs increase
with metallicity.  Again, this is best explained by higher
depletion levels at higher metallicity.
The GRB-DLAs do not exhibit this trend;  the values
are uniformly large.  This suggests
that the majority of GRB-DLAs have large dust-to-metal ratios.
Given the uncertainty on the nucleosynthesis of Zn, however, it is best
to address this issue from yet another angle.

In Figure~\ref{fig:relabnd}c we present upper limits to the Ti/Fe ratios
of those GRB-DLAs with observations of the Ti\,II~$\lambda 1910$ transitions.
In stellar atmospheres, Ti behaves like an $\alpha$-element
\citep[i.e., Ti tracks Si,O,Mg in 
metal-poor stars;][]{mcw97,pnc+00}.  In the Galactic ISM, Ti is 
highly refractory and one generally observes [Ti/Fe]$_{ISM} < 0$
\citep{jenkins87}.  Therefore, \cite{dpo02} emphasized that 
observations of 
Ti/Fe lend to degenerate-free interpretation:
super-solar Ti/Fe ratios indicate $\alpha$-enhancement whereas
sub-solar Ti/Fe ratios require differential depletion.
Therefore, the GRB-DLAs with [Ti/Fe]~$\lesssim 0$\,dex imply
substantial depletion, especially in light of the large
$\alpha$/Fe ratios.   
We consider this definitive evidence that at least some
GRB-DLAs are highly depleted.

The results in Figures~\ref{fig:relabnd}b,c argue that the GRB-DLAs
have at least modest depletion levels.  We contend
that differential depletion contributes at least +0.3\,dex to [$\alpha$/Fe]$_g$
for most GRB-DLAs, i.e.\ at least
50$\%$ or more of the Fe is locked into dust grains.
It remains an open question, therefore, whether the intrinsic
$\alpha$/Fe ratios of GRB-DLAs are enhanced relative to solar
as expected for young, star-forming regions \citep[e.g.][]{cdp07}.
While common practice is to examine the $\alpha$/Zn ratio (again, using Zn as a 
proxy for Fe), these results are especially
subject to the nucleosynthetic history of Zn.  Nevertheless, examining
$\alpha$/Zn we find that the GRB-DLAs 
primarily show lower limits that are consistent with the solar abundance
but allow for large enhancements.
In passing, we note that a large dataset of $\alpha$/Zn measurements for the
GRB-DLAs could inform the
processes of Zn production and also interpretations of these
measurements in QSO-DLAs \citep{nca+04}.

\subsubsection{N/$\alpha$}
\label{sec:nalpha}

Another abundance ratio of particular interest in terms
of nucleosynthesis is N/$\alpha$.
Because N is believed to be produced primarily in the AGB
phenomenon of intermediate mass stars \citep[e.g.][]{mm02}, the
N/$\alpha$ ratio provides a diagnostic of the star formation history
of the galaxy, especially at early times \citep{hek00}.
Atomic nitrogen is the dominant ionization state in neutral regions and,
unfortunately, this low-ion's transitions all
lie within the \lya\ forest. 
Therefore, the N abundance is difficult to measure in absorption-line
systems:  high resolution data is required and line-profile
analysis is often required to recover N$^0$ column densities.
Because N is non-refractory, the
observed ratios should nearly correspond to the intrinsic
values.  The only serious systematic uncertainties are ionization
corrections \citep{pho+02} which should be small
for GRB-DLAs given the very large \ion{H}{1} column 
densities.

The current GRB-DLA sample only allows one to 
constrain the N$^0$ column density for three GRB-DLAs:
GRB050730, GRB050820, GRB050922C.  The N/$\alpha$ ratios are presented in 
Figure~\ref{fig:relabnd}d in comparison with measurements made
for the QSO-DLAs \citep{pho+02,cmv+03,dz06}.  
Although this is a small sample, it is evident that there is a large
dispersion in [N/$\alpha$] for the GRB-DLAs.
This indicates that their host galaxies have experienced
a diverse range of star formation histories.

Two of the N/$\alpha$ values (GRB~050820,GRB~050730)
bracket the locus of observations
for the QSO-DLAs.  These values are relatively high ([N/$\alpha$]~$\ge -1$)
indicating a significant enhancement by intermediate mass stars.
This is not surprising for GRB~050820 where the gas metallicity
is large, implying several generations of star formation.
The N/$\alpha$ value for GRB~050730 (at metallicity [M/H]~$\approx -2$)
suggests an age older than 200\,Myr \citep{hek00},
which at $z=4$ implies a formation redshift approaching
the epoch of reionization.

The most intriguing measurement, however, is the upper limit 
to N/$\alpha$ in GRB~050922C.  This limit lies well below
even the so-called lower plateau of QSO-DLA values \citep{pho+02,cmv+03}.
If this upper limit has a nucleosynthetic origin, it implies
gas that has not been polluted by any intermediate mass stars.
Given this rather extreme value,
it is important to further address ionization corrections;
perhaps the low N/$\alpha$ value indicates an
unusual ionization state for the gas (i.e.\ a high
N$^+$/N$^0$ ratio).  Because atomic nitrogen has a relatively large cross-section
to X-rays \citep{sj98}, it may be under-abundant even in neutral
regions.  Both the GRB afterglow and the progenitor 
imply an enhanced radiation field which could lead to high
N$^+$/N$^0$ ratios.
We are especially concerned about photoionization for GRB~050922C
given that it has a low Mg$^0$ column density and likely 
absorption by the \ion{N}{5} doublet \citep{pwf+07}.
Equilibrium photoionization 
calculations, however, indicate ionization corrections are generally
less than 0.5\,dex, even for an input spectrum as hard as a 
quasar \citep{pro_ion02}.  In a future paper, we will perform
a time-dependent calculation to track the effects of the 
afterglow on this and other ionization states relevant to our
observations.   

Unless the ionization corrections are larger than 0.5\,dex, the limit
on [N/$\alpha$] lies below that of any other astrophysical environment.
This is not especially surprising for a GRB event, however, 
especially one with very low metallicity.  In this case, the
galaxy may have initiated star formation only very recently and
the observed metals may have no contribution from intermediate
mass stars.  
For a standard initial mass function, this requires an enrichment
age of less than $\sim 30$\,Myr \citep{hp07}.
Future observations will reveal whether low 
N/$\alpha$ values are characteristic of a larger sample of GRB-DLAs.

\begin{figure}[ht]
\begin{center}
\includegraphics[height=3.5in,angle=90]{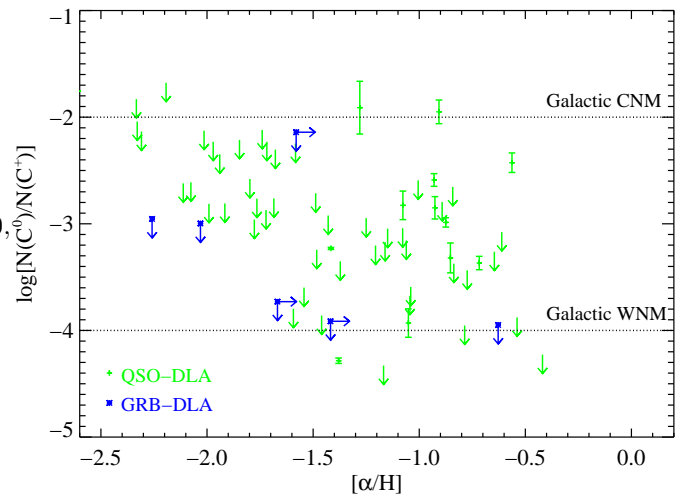}
\end{center}
\caption{C$^0$/C$^+$ ratios of the GRB-DLAs (dark) and QSO-DLAs (light)
inferred from column density measurements of \ion{C}{1} transitions
and by assuming $\log \N{C^+} = {\rm [\alpha/H]} + \log \mnhi - 4$.
Typical values of the cold neutral medium of the Galactic ISM are
${\rm C^0/C^+} = 10^{-2}$ \citep{js79} while significantly lower
values (10$^{-4}$) correspond to the warm neutral medium \citep{liszt02}.
We interpret the observations as more reflective of the latter phase
of the ISM, but caution that the results are sensitive to the far-UV
intensity, metallicity, and cosmic-ray flux through the gas.
}
\label{fig:ci}
\end{figure}

\subsection{Atomic Carbon}
\label{sec:ci}

In the Galaxy, an excellent tracer of cold, dense gas is atomic
carbon.  Because C$^0$ has an ionization potential below 1\,Ryd,
it is generally photoionized by the ambient far-UV radiation field
of the ISM.  In cold and dense regions, however, the recombination
rate (proportional to the electron and carbon densities)
is sufficiently large that cold clouds frequently show detectable
absorption from \ion{C}{1} transitions \citep[e.g.][]{jt01}.  For 
standard Galactic ISM conditions, the C$^0$/C$^+$ ratio is 
$\approx 10^{-2}$ in the cold neutral medium and decreases to 
$\approx 10^{-4}$ in the warm neutral medium \citep{liszt02}.
These ratios, of course, are sensitive to the gas metallicity,
far-UV radiation field, and cosmic ray flux of the ISM
\citep{liszt02,wpg03}.  Nevertheless, very low C$^0$/C$^+$ values
generally imply gas with density less than 10 particles per cm$^3$.

Figure~\ref{fig:ci} presents the C$^0$/C$^+$ ratios estimated
for the GRB-DLAs and QSO-DLAs (sample HR-A).  While we can measure 
$\N{C^0}$ directly from \ion{C}{1} transitions, the \ion{C}{2} transitions
are generally too saturated to provide even a valuable lower limit
to $\N{C^+}$.  Therefore, we infer $\N{C^+}$ from the [$\alpha$/H] abundance
and \nhi\ value, assuming [$\alpha$/C]=+0.3 
\citep[i.e.\ a modest $\alpha$-enhancement;][]{acn+04}: 

\begin{equation}
\log \N{C^+} = {\rm [\alpha/H]} + \log \mnhi - 12 + 8.3 \;\;\; .
\end{equation}
Note that this calculation is actually independent of the \nhi\ value.
None of the GRB-DLAs in our sample has a positive detection, although
\cite{vls+07} report the detection of \ion{C}{1} in GRB-DLA~060418
with a value consistent with our upper limit.
Because the metal-line column densities of the GRB-DLAs are large,
the upper limits to C$^0$/C$^+$ generally lie below those
of the QSO-DLAs even though the latter are derived from higher S/N data.
The observations suggest the gas in GRB-DLAs (and many QSO-DLAs)
is dominated by a warm, less dense phase.  We caution again, however,
that the predicted C$^0$/C$^+$ values are sensitive to the 
far-UV flux, which is likely to be enhanced near GRB.
A more cautious interpretation of the data is that it offers
additional, yet non-conclusive support that GRB-DLA gas
is not characteristic of the Galactic cold neutral medium.

\section{DISCUSSION}
\label{sec:discuss}

\subsection{General Characteristics of the ISM Near Star Forming Regions}
\label{sec:grbdiscuss}

Reviewing the results presented in the previous section, we 
summarize the properties of gas observed surrounding GRB afterglows
as having:

\begin{itemize}

\item Very large \ion{H}{1} column densities with median 
$\mnhi \approx 10^{21.6} \cm{-2}$.

\item A wide range of metallicities (1/100 to nearly solar)
with median larger than 1/10 solar. 

\item Observed $\alpha$/Fe ratios in excess of $3\times$ the solar abundance
reflecting enrichment by massive stars (Type\,II SN) and/or differential
depletion.

\item Large [Zn/Fe] values and several solar or sub-solar 
[Ti/Fe] upper limits which indicate substantial differential depletion.

\item A dispersion of [N/$\alpha$] indicating a diverse
set of star formation histories.   We identify one case
(GRB~050922C) with an extremely low value 
suggesting gas enriched solely by massive stars.

\end{itemize}

At first glance, the observation of large \ion{H}{1} surface densities
and significantly depleted, chemically enriched gas appear to reflect
the conditions expected for actively star-forming regions (i.e.\ molecular
clouds).  Quantitatively, however, we find that the GRB-DLA 
properties are more characteristic of the ambient ISM of modern galaxies.
As several authors have noted for individual GRB-DLA \citep{sf04,pbf+06,sbp+06},
the observed depletion levels are more representative of warm (i.e.\ less dense)
clouds in the Galactic ISM \citep[see also][]{savaglio06}.
On the other hand, the majority of GRB-DLAs have sub-solar metallicity
and a direct comparison with the Galactic ISM may not be appropriate.
Unfortunately, the differential depletion characteristics of  
cold, dense gas in other galaxies (e.g.\ the Magellanic clouds) has
not been extensively measured,
in part because dust obscuration challenges the analysis.
In the Small Magellanic Cloud (SMC), one does observe highly 
depleted gas along the sightline to Sk~155 
but one also finds modest depletions in clouds where \ion{C}{1}
analysis indicates large 
density \citep[$>100 \cm{-3}$][]{welty01,sgc+06,welty06}.
We cautiously conclude
that the depletion levels observed for the GRB-DLAs
are generally lower than that observed for 
cold, dense clouds in the local universe.
It would be very valuable to independently estimate the density of gas
responsible for GRB-DLAs.

Throughout this paper, we have presented additional
observations which argue the
GRB-DLAs do not arise in a cold, dense phase. First, the
gas has very low molecular fraction \citep[][Paper~III]{vel+04}. 
Although a low molecular fraction could result from photoionization
by the progenitor, nearby O and B stars, and the GRB afterglow,
we have argued the gas lies at $\gtrsim 100$\,pc from the GRB
where these effects may be lessened.
If the gas has large particle density ($n_H > 10^{3} \cm{-3}$),
then it would be difficult to maintain such a low H$_2$ fraction 
(Paper~III). 

Second, the only atomic line routinely measured
in the GRB-DLAs is \ion{Mg}{1} and its large dielectronic recombination
coefficient does not require cold, dense gas.  In $\S$~\ref{sec:ci}
we presented upper limits on C$^0$/C$^+$ which argue against gas
with characteristics like the dense CNM of the Milky Way.
Third, there are no especially anomalous abundance patterns observed
for the gas (e.g.\ extreme metallicity, very high $\alpha$/Fe ratios)
which could occur if the gas represented partially mixed SN ejecta
or circumstellar material.
Altogether, we conclude that the gas revealed by GRB afterglow
spectra represents the ISM near the GRB but not gas directly
associated with its own molecular cloud. 

\subsection{Comparisons with QSO-DLAs}

In the previous subsection, we argued that the GRB-DLAs represent
gas from the ambient ISM of their host galaxies.  The large \nhi\
values of QSO-DLAs argue these sightlines also penetrate the ISM
of high $z$ galaxies.   Furthermore, the majority of QSO-DLAs have
depletion patterns, low C$^+0$/C$^+$ ratios, and low molecular
fractions which are not characteristic of cold, dense gas in the Galaxy.
In this respect, it is fair to compare the observations of the two
populations to address whether they are drawn from the same parent
population of galaxies.
In $\S$~\ref{sec:gas} we compared the \nhi\ values, metallicity,
and relative abundance ratios of the QSO-DLAs and GRB-DLAs.
The latter are characterized by higher \nhi\ values, higher
$\alpha$/Fe and Zn/Fe ratios, and higher metallicities. 
These differences beg the question:
{\it Are GRB-DLAs and QSO-DLAs drawn from distinct 
distributions of host galaxies?}

We contend that the observations presented in this paper do not 
contradict the null hypothesis that GRB-DLAs and QSO-DLAs have
identical parent populations of galaxies but their differences
can be explained by sightlines with distinct impact parameter 
distributions.
Although offsets exist in [M/H] and the X/Fe ratios, 
the differences are relatively small ($<0.5$\,dex).
Typical observed gradients in the metallicity of 
stars and \ion{H}{2} regions of local galaxies imply a variation of 
at least 0.3\,dex from the center of the galaxy to the radii roughly
corresponding to $\mnhi = 2\sci{20} \cm{-2}$ \citep[e.g.][]{kbg03,ckr05,lka06}.  
Furthermore, the \ion{H}{1} surface density and volume density are undoubtedly
larger toward the center of the galaxy (at least until the gas 
becomes molecular).  And it is reasonable to expect that higher
metallicity, denser gas will exhibit larger depletion levels than
more tenuous gas at the outer edges of \ion{H}{1} disks.

At present, the only distribution with an especially large
offset ($>1$\,dex difference in median value)
is the \nhi\ values.
While differences in the impact parameters of GRB-DLAs and QSO-DLAs
may explain the offsets in the median of the distributions for
\nhi, [M/H], and X/Fe ratios, the identification of many GRB-DLAs
with $\mnhi > 10^{22} \cm{-2}$ and {\it none} in the QSO-DLA sample
is striking.  
In $\S$~\ref{sec:hgas}, we demonstrated that the GRB-DLA \nhi\ distribution
is reasonably well explained by an \ion{H}{2} region embedded with
an exponential \ion{H}{1} disk with central column 
density $\log N_0 = 10^{22} \cm{-2}$.  
While this simplistic model is a good description of the
observations, it does lead to a conflict with the \ion{H}{1}
frequency distribution \fnhi\ observed for the QSO-DLAs.
If the majority of high $z$ galaxies have $\log N_0 = 10^{22} \cm{-2}$,
then one would not observe a break 
in \fnhi\ at $\mnhi \approx 10^{21.5} \cm{-2}$
as observed \citep{phw05}.   One possible resolution for this
conflict is that GRB-DLA sightlines are obscured by dust from magnitude-limited
QSO surveys.  
Let us now consider this hypothesis in greater detail.

\subsection{Can QSO-DLA Samples Probe GRB-DLA Sightlines?}

As Figure~\ref{fig:nhihist} reveals, a significant fraction of GRB-DLAs
have $\mnhi \geq 10^{22} \cm{-2}$ yet not one of the $\approx 500$
QSO-DLAs discovered thus far have $\mnhi > 10^{21.9} \cm{-2}$.
An analysis of the SDSS dataset actually requires that the QSO-DLAs
exhibit a break in their \ion{H}{1} frequency distribution \fnhi\
at $\mnhi \approx 10^{21.5} \cm{-2}$ to explain the absence
of large \nhi\ absorbers \citep{phw05}.  This is most
easily seen by extrapolating the $\mfnhi \propto \mnhi^{-2}$ power-law
observed at $\mnhi < 10^{21} \cm{-2}$ to larger column densities.
One predicts that there should be $\approx 10$ damped \lya\ systems
with $\mnhi \geq 10^{22} \cm{-2}$ per 500 observed.
The absence of a single QSO-DLAs with this column density 
indicates that \fnhi\ has a steeper than $\mnhi^{-2}$ dependence
at these column densities.  
The observation of a break in \fnhi\ 
and the routine observation of $\mnhi > 10^{22} \cm{-2}$ in 
GRB-DLAs raises concerns that dust obscuration is biasing the
QSO-DLA sample against large \nhi\ values.

There is additional (circumstantial)
evidence for a dust obscuration bias in QSO-DLAs.
First, QSO-DLAs with a combination of large \nhi\ and high 
metallicity (metal-strong DLA), i.e.\ potentially large dust columns,
are rare \citep{blb+98,shf06}.  
Second, every DLA is enriched by heavy metals with metallicity 
[M/H]~$> -3$ \citep{pgw+03} and many have relative abundance
ratios (e.g.\ Zn/Fe) indicative of differential dust depletion
\citep{pettini94,pw02}.  
Third, several quasars with intervening DLA show significant
reddening that can be attributed to the absorber \cite{vcl+06}.
Motivated by these observations, we will test whether
QSO-DLA samples would be likely to include GRB-DLA
sightlines if background quasars were placed directly on the
sightline.  

We begin by deriving the dust-to-gas ratios of the GRB-DLAs. 
In the following, we will assume the GRB-DLA dust properties
are best matched by those observed for the SMC.  We make this choice
because 
(i) no GRB-DLA to date has exhibited the 2175\AA\ dust feature
characteristic of the Milky Way and LMC extinction 
laws \citep[e.g.][]{mhk+02,sf04,evl+06,bp+07};
and 
(ii) the metallicities of GRB-DLAs are representative of the SMC.
We derive the dust-to-gas ratio of the DLAs relative 
to the SMC: $\kappa/\kappa_{SMC}$ by assuming the SMC has metallicity 
[M/H]$_{SMC} = -0.7$ and that 90$\%$ of its refractory
elements are depleted from the gas-phase, 
i.e.\ [M/Fe]$_{SMC} = +1$ \citep{welty97,welty01}. 
In this case,  the relative dust-to-gas ratio is
\begin{equation}
\frac{\kappa}{\kappa_{SMC}} = 10^{ {\rm [M/H]} + 0.7} 
\ltk \frac{1 - 10^{(\Delta_i - {\rm [M/Fe]})}}{1 - 10^{-1}} \rtk
\end{equation}
where $\Delta_i$ allows for a nucleosynthetic contribution to the observed
[M/Fe] value and we assume $\Delta_i = 0$ for the SMC. 
For the GRB-DLAs, we adopt the [M/H] and [M/Fe] values listed in 
Table~\ref{tab:abnd}.
In the majority of cases, M is sulfur and the remainder
assume Zn or Si.
In the following, we will assume $\Delta_i = 0.3$\,dex for the GRB-DLAs
and 0.2\,dex for the QSO-DLAs.  
The high specific star formation rates observed for GRB host galaxies
indicate young ages and, therefore, gas enriched by massive stars
giving $\alpha$-enriched abundances ([$\alpha$/Fe]~$>+0.2$).
Similarly, the QSO-DLAs show a plateau of [Si/Fe]~=+0.25\,dex 
values at low metallicity indicative of Type\,II enrichment \citep{pw02}, and
enhanced $\alpha$/Fe ratios in ``dust free'' systems \citep{dz06}.
In any case, the results are not sensitive to
our choices for $\Delta_i$ unless we allow $\Delta_i > 0.5$\,dex 
for the GRB-DLAs
or $\Delta_i > 0.3$\,dex for the QSO-DLAs. 

\begin{figure}[ht]
\begin{center}
\includegraphics[height=3.5in,angle=90]{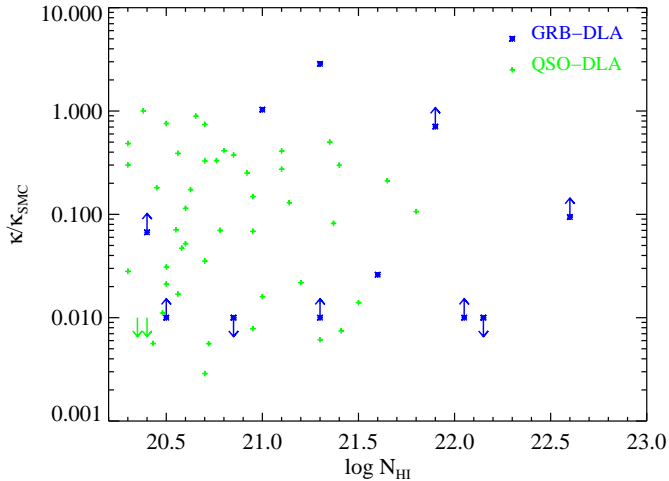}
\end{center}
\caption{Dust-to-gas ratio $\kappa$ of the GRB-DLAs (dark blue) and
QSO-DLAs (light green) relative to the dust-to-gas ratio of the SMC
($\kappa_{SMC}$)
as a function of \ion{H}{1} column density.
}
\label{fig:dusttogas}
\end{figure}

Figure~\ref{fig:dusttogas} shows the $\kappa/\kappa_{SMC}$ values for
the GRB-DLAs and also the echelle sample (HR-E) of QSO-DLAs.  Both samples
exhibit a wide range of dust-to-gas ratios relative to the SMC.
We also find that 
the GRB-DLAs show a higher fraction of large values
($\kappa/\kappa_{SMC} \gtrsim 1$).  
Using the observed \nhi\ values and assuming SMC dust properties, we
can convert these dust-to-gas ratios into visual extinction, $A_V$.
Specifically, we adopt two empirical relations:  (i) the observed
reddening in the SMC \citep{tsr+02}
\begin{equation}
\log E(B-V) = \log \mnhi - 22.95
\label{eqn:ebv}
\end{equation}
and (ii) a linear relation between $A_V$ and $E(B-V)$ \citep{gcm+03}
modified by the dust-to-gas ratio
\begin{equation}
A_V = 2.74 \; E(B-V) \; \frac{\kappa}{\kappa_{SMC}} \;\; .
\label{eqn:av}
\end{equation}

\begin{figure}[ht]
\begin{center}
\includegraphics[height=3.5in,angle=90]{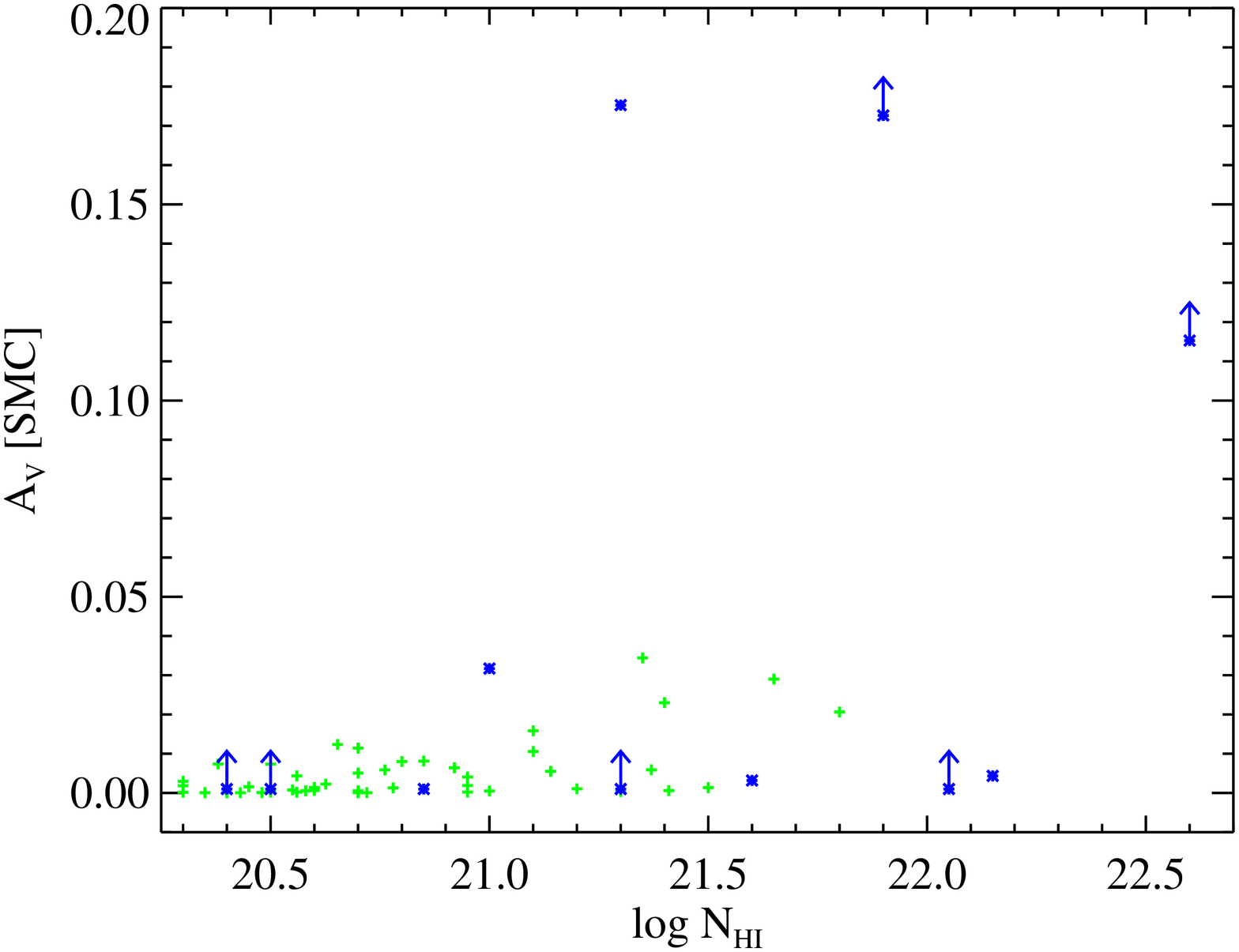}
\end{center}
\caption{Predicted visual extinctions $A_V$ for the GRB-DLA and
QSO-DLA sightlines assuming the dust-to-gas ratios presented in
Figure~\ref{fig:dusttogas} and assuming the dust properties of the SMC
\citep{tsr+02,gcm+03}.
All of the QSO-DLA sightlines show very small values ($A_V < 0.05$\,mag).
In contrast, the GRB-DLA show a nearly bimodal population with 
several small values and several sightlines
approaching 0.2\,mag.  Note that
$A_V=0.1$\,mag corresponds to an extinction of 0.6\,mag assuming the SMC
extinction law. 
}
\label{fig:smcav}
\end{figure}

Figure~\ref{fig:smcav} presents the $A_V$ values against \ion{H}{1}
column density for the GRB-DLAs compared against QSO-DLAs.  All of
the QSO-DLAs have $A_V < 0.05$\,mag \citep[see also][]{pw02}.
It is not surprising, therefore, that one observes very little 
reddening of quasar spectra by QSO-DLAs \citep{ml04}.
In contrast, the estimated $A_V$ values for approximately half the
GRB-DLAs are substantially larger than the QSO-DLA distribution
($A_V > 0.1$\,mag).  At $A_V=0.1$\,mag, the extinction at the
\lya\ transition is $A_{Ly\alpha} = 0.6$\,mag assuming the 
SMC extinction law \citep{prevot84}.  This is a considerable but
not overwhelming level of extinction.   
Interestingly, there is no obvious correlation between $A_V$ and
\nhi, at least for $A_V$ values greater than 0.05\,mag.
We also note that the $A_V$ 
values presented in Figure~\ref{fig:smcav} are roughly 
consistent with the $A_V$ values inferred from broad-band
photometry or spectrophotometry of GRB afterglows 
\citep[e.g.][]{mhk+02,vel+04,sf04,bph+06}.
On the other hand, our $A_V$ values appear to be at odds
with those inferred from X-ray absorption measurements
\citep[e.g.][]{wfl+06,bph+06} which has led some authors
to invoke ``grey'' extinction laws for the dust enveloping
GRBs \citep{gw01}.
This issue will be addressed in greater detail in a 
future paper (Butler et al.\ 2007, in prep).

\begin{figure}[ht]
\begin{center}
\includegraphics[height=3.5in,angle=90]{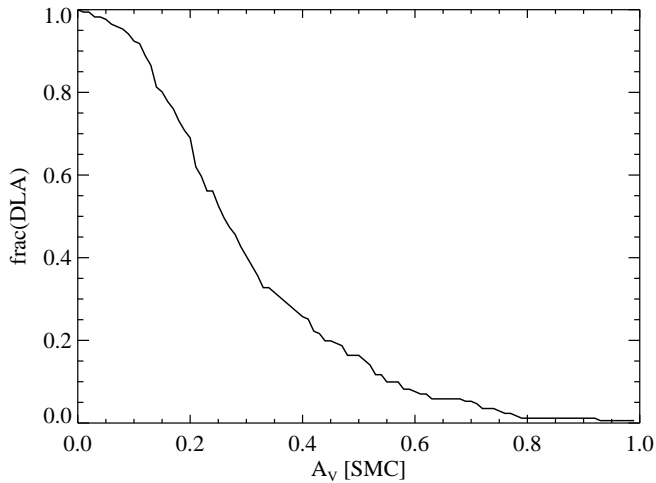}
\end{center}
\caption{Fraction of DLAs recovered from the SDSS QSO-DLA
survey of \cite{phw05} as a function of additional
visual extinction ($A_V$) for the DLAs.
For $A_V < 0.2$\,mag, the effect of obscuration
is mild while only a small fraction of QSO-DLAs 
with $A_V > 0.4$\,mag would be recovered in the SDSS
survey.
}
\label{fig:avdla}
\end{figure}

Finally, we examine whether the inferred $A_V$ values for the GRB-DLA
sightlines are sufficiently large to bias against their detection in 
QSO samples. 
To estimate the effects,
we assessed the signal-to-noise ratio (S/N) at the center of the
\lya\ profile of each QSO-DLAs in the SDSS-DR3 sample.  Specifically,
we measured the median S/N using the continuum adopted in the fits to the
\lya\ profiles by \cite{phw05}.  By the definition of their 
statistical sample, the S/N exceeds $4 \, {\rm pix^{-1}}$ in each case.
We then lowered the S/N by adopting an extinction $A_V$ assuming
that the noise is quasar dominated,
\begin{equation}
{\rm S/N} = ({\rm S/N}_0) / \sqrt{10^{A_{Ly\alpha}/2.5}}
\end{equation}
The fraction of QSO-DLAs that satisfy the S/N$>4 \, {\rm pix^{-1}}$ criterion
is shown in Figure~\ref{fig:avdla} as a function of $A_V$.
For $A_V < 0.2$\,mag, dust obscuration has a relatively small
impact on the QSO-DLA sample.  We cautiously conclude, therefore,
that the majority of observed GRB-DLAs would also be observed along quasar
sightlines.

This conclusion, however, is subject to a few caveats.  First,
the exercise described by Figure~\ref{fig:avdla} does not account for 
quasars that would become too red or too faint to satisfy the 
QSO target criteria for the SDSS \citep{rng+04}.
We suspect, however, that this effect is much smaller than the S/N cut
we have applied (Murphy et al.\ 2007, in prep).
Second, several of the $\kappa/\kappa_{SMC}$ values for the GRB-DLAs are
lower limits because of line saturation.  It is possible that these 
have $A_V > 0.4$\,mag and would have a small probability of entering
the QSO-DLA sample.  Third, all of the results assume SMC extinction.
This assumption has two implications:
(i) it establishes the reddening per \ion{H}{1} atom (scaled by metallicity;
Equation~\ref{eqn:ebv}) and 
(ii) the shape of the extinction law which determines $A_{Ly\alpha}/A_V$.
Regarding the latter aspect, the SMC has the steepest extinction
law of local galaxies.  Other (``grayer'') dust
laws would generally lead to less extinction at \lya.
The reddening per \ion{H}{1} atom in the SMC, however, is smaller than
the Milky Way and not only because of the SMC's lower metallicity 
\citep{welty06}. 
Fourth, we have ignored the possibility that a significant column
of dust exists in the \ion{H}{2} region surrounding the GRB.
This could be revealed by broad-band photometry of the afterglow
compared against $A_V$ estimations from the observed metal column
densities.  Finally, we note that a non-negligible fraction of 
GRB afterglows are severely reddened, most likely by dust
surrounding the event \citep{lfr+06,pdl+06}.
These ``dark bursts'' do not enter our GRB-DLA sample and
likely have $A_V$ values much larger than those presented in 
Figure~\ref{fig:smcav}.  At present, this extremely dusty gas
is not well probed by GRB or QSO observations.

These concerns aside,
our analysis suggests that many (if not all) of the GRB-DLAs in our
sample
could be detected in modern QSO-DLA surveys.  If confirmed by larger
samples, then the very low incidence of QSO-DLAs 
with $\mnhi \geq 10^{22} \cm{-2}$ must be explained by a substantial
decrease in the cross-section of gas with these column densities.
To avoid conflict with the observed \fnhi\ distribution, one requires that
the cross-section of the regions probed by GRB-DLAs must be
less than $1\%$ of the cross-section corresponding to the damped \lya\
criterion, $\log \mnhi = 20.3$.  Characterizing the 
latter by a radius $r_{20.3}$, we require the $r_{GRB} < r_{20.3}/10$.
This relation is rather easily achieved if $r_{GRB}$ is 100pc or less,
but could pose a problem if $r_{GRB} \gtrsim 1$\,kpc \citep{vls+07}.

\acknowledgments

We wish to thank Arthur M. Wolfe for helpful discussions and
for allowing us to analyze
results on QSO-DLAs prior to the public release of these data.
We thank E. Ramirez-Ruiz for constructive comments on an early
version of this manuscript.
J.X.P. is partially supported by NASA/Swift grant NNG05GF55G
and an NSF CAREER grant (AST-0548180).


\begin{thebibliography}{}

\bibitem[{Akerman} {\it et al.}\ (2004)]{acn+04}
{Akerman}, C.~J. {\it et al.}\  2004, \aap, 414, 931.

\bibitem[{Barth} {\it et al.}\ (2003)]{bsc+03}
{Barth}, A.~J. {\it et al.}\  2003, ApJ ({\it Letters}), 584, L47.

\bibitem[{Berger} {\it et al.}\ (2006)]{bpck+05}
{Berger}, E. {\it et al.}\  2006, \apj, 642, 979.

\bibitem[{Blitz}(1993)]{blitz93}
{Blitz}, L. 1993, in { Protostars and Planets III}, ed.\ E.~H. {Levy} and J.~I.
  {Lunine}, 125.

\bibitem[{Blitz} {\it et al.}\ (2007)]{bfk+07}
{Blitz}, L. {\it et al.}\  2007, in { Protostars and Planets V}, ed.\ B.
  {Reipurth}, D. {Jewitt}, and K. {Keil}, 81.

\bibitem[{Bloom}, {Kulkarni} \& {Djorgovski}(2002)]{bkd02}
{Bloom}, J.~S., {Kulkarni}, S.~R., and {Djorgovski}, S.~G. 2002, AJ, 123, 1111.

\bibitem[{Boisse} {\it et al.}\ (1998)]{blb+98}
{Boisse}, P. {\it et al.}\  1998, \aap, 333, 841.

\bibitem[{Butler} {\it et al.}\ (2007)]{bp+07}
{Butler}, N. {\it et al.}\  2007, In prep.

\bibitem[{Butler} {\it et al.}\ (2006)]{bph+06}
{Butler}, N.~R. {\it et al.}\  2006, \apj, 652, 1390.

\bibitem[{Calura}, {Dessauges-Zavadsky} \& {Prochaska}(2007)]{cdp07}
{Calura}, F., {Dessauges-Zavadsky}, M., and {Prochaska}, J. 2007, In prep.

\bibitem[{Calura}, {Matteucci} \& {Vladilo}(2003)]{cmv03}
{Calura}, F., {Matteucci}, F., and {Vladilo}, G. 2003, \mnras, 340, 59.

\bibitem[{Castro} {\it et al.}\ (2003)]{cgh+03}
{Castro}, S. {\it et al.}\  2003, ApJ, 586, 128.

\bibitem[{Centuri{\'o}n} {\it et al.}\ (2003)]{cmv+03}
{Centuri{\'o}n}, M. {\it et al.}\  2003, \aap, 403, 55.

\bibitem[{Chen}, {Kennicutt} \& {Rauch}(2005)]{ckr05}
{Chen}, H.-W., {Kennicutt}, Jr., R.~C., and {Rauch}, M. 2005, \apj, 620, 703.

\bibitem[{Chen} {\it et al.}\ (2007)]{cpb+07}
{Chen}, H.-W. {\it et al.}\  2007, ArXiv Astrophysics e-prints.

\bibitem[{Chen} {\it et al.}\ (2005)]{cpb+05}
{Chen}, H.-W. {\it et al.}\  2005, \apjl, 634, L25.

\bibitem[{Christensen}, {Hjorth} \& {Gorosabel}(2004)]{chg04}
{Christensen}, L., {Hjorth}, J., and {Gorosabel}, J. 2004, A\&A, 425, 913.

\bibitem[{Dekker} {\it et al.}\ (2000)]{uves}
{Dekker}, H. {\it et al.}\  2000, in { Proc. SPIE Vol. 4008, p. 534-545,
  Optical and IR Telescope Instrumentation and Detectors, Masanori Iye; Alan F.
  Moorwood; Eds.}, ed.\ M. {Iye} and A.~F. {Moorwood}, 534.

\bibitem[{Dessauges-Zavadsky} {\it et al.}\ (2007)]{dz07}
{Dessauges-Zavadsky}, M. {\it et al.}\  2007, \aap.

\bibitem[{Dessauges-Zavadsky} {\it et al.}\ (2006)a]{dcp+06}
{Dessauges-Zavadsky}, M. {\it et al.}\  2006a, \apjl, 648, L89.

\bibitem[{Dessauges-Zavadsky} {\it et al.}\ (2001)]{mirka01}
{Dessauges-Zavadsky}, M. {\it et al.}\  2001, \aap, 370, 426.

\bibitem[{Dessauges-Zavadsky}, {Prochaska} \& {D'Odorico}(2002)]{dpo02}
{Dessauges-Zavadsky}, M., {Prochaska}, J.~X., and {D'Odorico}, S. 2002, \aap,
  391, 801.

\bibitem[{Dessauges-Zavadsky} {\it et al.}\ (2006)b]{dz06}
{Dessauges-Zavadsky}, M. {\it et al.}\  2006b, \aap, 445, 93.

\bibitem[{Draine} \& {Hao}(2002)]{draine02}
{Draine}, B.~T. and {Hao}, L. 2002, \apj, 569, 780.

\bibitem[{Ellison} {\it et al.}\ (2006)]{evl+06}
{Ellison}, S.~L. {\it et al.}\  2006, \mnras, 372, L38.

\bibitem[{Fall} \& {Pei}(1993)]{fall93}
{Fall}, S.~M. and {Pei}, Y.~C. 1993, \apj, 402, 479.

\bibitem[{Feigelson} \& {Nelson}(1985)]{fn85}
{Feigelson}, E.~D. and {Nelson}, P.~I. 1985, \apj, 293, 192.

\bibitem[{Fenner}, {Prochaska} \& {Gibson}(2004)]{fpg04}
{Fenner}, Y., {Prochaska}, J.~X., and {Gibson}, B.~K. 2004, \apj, 606, 116.

\bibitem[{Fiore} {\it et al.}\ (2005)]{fdl+05}
{Fiore}, F. {\it et al.}\  2005, \apj, 624, 853.

\bibitem[{Fruchter}, {Krolik} \& {Rhoads}(2001)]{fkr01}
{Fruchter}, A., {Krolik}, J.~H., and {Rhoads}, J.~E. 2001, \apj, 563, 597.

\bibitem[{Fruchter} {\it et al.}\ (2006)]{fls+06}
{Fruchter}, A.~S. {\it et al.}\  2006, \nat, 441, 463.

\bibitem[{Fynbo} {\it et al.}\ (2006)a]{fsl+06}
{Fynbo}, J.~P.~U. {\it et al.}\  2006a, \aap, 451, L47.

\bibitem[{Fynbo} {\it et al.}\ (2006)b]{fwt+06}
{Fynbo}, J.~P.~U. {\it et al.}\  2006b, \nat, 444, 1047.

\bibitem[{Galama} \& {Wijers}(2001)]{gw01}
{Galama}, T.~J. and {Wijers}, R.~A.~M.~J. 2001, ApJ ({\it Letters}), 549, L209.

\bibitem[{Gehrels}(2000)]{geh00}
{Gehrels}, N.~A. 2000, in { Proc. SPIE X-Ray and Gamma-Ray Instrumentation for
  Astronomy XI}, ed.\ Kathryn~A. Flanagan and Oswald~H. Siegmund, volume 4140,
  42.

\bibitem[{Gordon} {\it et al.}\ (2003)]{gcm+03}
{Gordon}, K.~D. {\it et al.}\  2003, \apj, 594, 279.

\bibitem[{Guetta} \& {Piran}(2007)]{gp07}
{Guetta}, D. and {Piran}, T. 2007, ArXiv Astrophysics e-prints.

\bibitem[{Hammer} {\it et al.}\ (2006)]{hfs+06}
{Hammer}, F. {\it et al.}\  2006, \aap, 454, 103.

\bibitem[{Henry} \& {Prochaska}(2007)]{hp07}
{Henry}, D. and {Prochaska}, J. 2007, In prep.

\bibitem[{Henry}, {Edmunds} \& {K{\"o}ppen}(2000)]{hek00}
{Henry}, R.~B.~C., {Edmunds}, M.~G., and {K{\"o}ppen}, J. 2000, \apj, 541, 660.

\bibitem[{Herbert-Fort} {\it et al.}\ (2006)]{shf06}
{Herbert-Fort}, S. {\it et al.}\  2006, \pasp, 118, 1077.

\bibitem[{Hjorth} {\it et al.}\ (2003)]{hsm+03}
{Hjorth}, J. {\it et al.}\  2003, Nature, 423, 847.

\bibitem[{Hoffman} {\it et al.}\ (1996)]{hwf+96}
{Hoffman}, R.~D. {\it et al.}\  1996, \apj, 460, 478.

\bibitem[{Howk} \& {Sembach}(1999)]{howk_ion99}
{Howk}, J.~C. and {Sembach}, K.~R. 1999, \apjl, 523, L141.

\bibitem[{Jakobsson} {\it et al.}\ (2006)a]{jfl+06}
{Jakobsson}, P. {\it et al.}\  2006a, \aap, 460, L13.

\bibitem[{Jakobsson} {\it et al.}\ (2006)b]{jlf+06}
{Jakobsson}, P. {\it et al.}\  2006b, \aap, 447, 897.

\bibitem[{Jenkins}(1986)]{jenkins86}
{Jenkins}, E.~B. 1986, \apj, 304, 739.

\bibitem[{Jenkins}(1987)]{jenkins87}
{Jenkins}, E.~B. 1987, in { ASSL Vol. 134: Interstellar Processes}, ed.\ D.~J.
  {Hollenbach} and H.~A. {Thronson}, Jr., 533.

\bibitem[{Jenkins} \& {Shaya}(1979)]{js79}
{Jenkins}, E.~B. and {Shaya}, E.~J. 1979, \apj, 231, 55.

\bibitem[{Jenkins} \& {Tripp}(2001)]{jt01}
{Jenkins}, E.~B. and {Tripp}, T.~M. 2001, \apjs, 137, 297.

\bibitem[{Kawai} {\it et al.}\ (2006)]{kka+06}
{Kawai}, N. {\it et al.}\  2006, \nat, 440, 184.

\bibitem[{Kennicutt}, {Bresolin} \& {Garnett}(2003)]{kbg03}
{Kennicutt}, Jr., R.~C., {Bresolin}, F., and {Garnett}, D.~R. 2003, \apj, 591,
  801.

\bibitem[{Kewley} {\it et al.}\ (2007)]{kbg+07}
{Kewley}, L.~J. {\it et al.}\  2007, \aj, 133, 882.

\bibitem[{Kobulnicky} \& {Kewley}(2004)]{kk04}
{Kobulnicky}, H.~A. and {Kewley}, L.~J. 2004, \apj, 617, 240.

\bibitem[{Langer} \& {Norman}(2006)]{ln06}
{Langer}, N. and {Norman}, C.~A. 2006, \apjl, 638, L63.

\bibitem[{Lanzetta}(1993)]{lzt93}
{Lanzetta}, K.~M. 1993, in { ASSL Vol. 188: The Environment and Evolution of
  Galaxies}, ed.\ J.~M. {Shull} and H.~A. {Thronson}, 237.

\bibitem[{Le Floc'h} {\it et al.}\ (2003)]{ldm+03}
{Le Floc'h}, E. {\it et al.}\  2003, A\&A, 400, 499.

\bibitem[{Ledoux} {\it et al.}\ (2006)]{ledoux06}
{Ledoux}, C. {\it et al.}\  2006, \aap, 457, 71.

\bibitem[{Ledoux}, {Petitjean} \& {Srianand}(2003)]{ledoux03}
{Ledoux}, C., {Petitjean}, P., and {Srianand}, R. 2003, \mnras, 346, 209.

\bibitem[{Levan} {\it et al.}\ (2006)]{lfr+06}
{Levan}, A. {\it et al.}\  2006, \apj, 647, 471.

\bibitem[{Liszt}(2002)]{liszt02}
{Liszt}, H. 2002, \aap, 389, 393.

\bibitem[{Lu} {\it et al.}\ (1996)]{lu96}
{Lu}, L. {\it et al.}\  1996, \apjs, 107, 475.

\bibitem[{Luck}, {Kovtyukh} \& {Andrievsky}(2006)]{lka06}
{Luck}, R.~E., {Kovtyukh}, V.~V., and {Andrievsky}, S.~M. 2006, \aj, 132, 902.

\bibitem[{McWilliam}(1997)]{mcw97}
{McWilliam}, A. 1997, \araa, 35, 503.

\bibitem[Metzger {\it et al.}\ (1997)]{mdk+97}
Metzger, M.~R. {\it et al.}\  1997, Nature, 387, 879.

\bibitem[{Meyer} \& {Roth}(1990)]{mr90}
{Meyer}, D.~M. and {Roth}, K.~C. 1990, \apj, 363, 57.

\bibitem[{Meynet} \& {Maeder}(2002)]{mm02}
{Meynet}, G. and {Maeder}, A. 2002, \aap, 381, L25.

\bibitem[{Mirabal} {\it et al.}\ (2006)]{mha+06}
{Mirabal}, N. {\it et al.}\  2006, \apjl, 643, L99.

\bibitem[{Mirabal} {\it et al.}\ (2002)]{mhk+02}
{Mirabal}, N. {\it et al.}\  2002, ApJ, 578, 818.

\bibitem[{Mo}, {Mao} \& {White}(1998)]{mmw98}
{Mo}, H.~J., {Mao}, S., and {White}, S.~D.~M. 1998, \mnras, 295, 319.

\bibitem[{Modjaz} {\it et al.}\ (2007)]{mkk+07}
{Modjaz}, M. {\it et al.}\  2007, ArXiv Astrophysics e-prints.

\bibitem[{Molaro} {\it et al.}\ (2001)]{molaro01}
{Molaro}, P. {\it et al.}\  2001, \apj, 549, 90.

\bibitem[{Murphy} \& {Liske}(2004)]{ml04}
{Murphy}, M.~T. and {Liske}, J. 2004, \mnras, 354, L31.

\bibitem[{Nissen} {\it et al.}\ (2004)]{nca+04}
{Nissen}, P.~E. {\it et al.}\  2004, \aap, 415, 993.

\bibitem[{Pellizza} {\it et al.}\ (2006)]{pdl+06}
{Pellizza}, L.~J. {\it et al.}\  2006, \aap, 459, L5.

\bibitem[{Penprase} {\it et al.}\ (2006)]{pbf+06}
{Penprase}, B.~E. {\it et al.}\  2006, \apj, 646, 358.

\bibitem[{Perna} \& {Lazzati}(2002)]{pl02}
{Perna}, R. and {Lazzati}, D. 2002, \apj, 580, 261.

\bibitem[{Pettini} {\it et al.}\ (1994)]{pettini94}
{Pettini}, M. {\it et al.}\  1994, \apj, 426, 79.

\bibitem[{Piranomonte} {\it et al.}\ (2007)]{pwf+07}
{Piranomonte}, S. {\it et al.}\  2007, \aap.

\bibitem[{Prevot} {\it et al.}\ (1984)]{prevot84}
{Prevot}, M.~L. {\it et al.}\  1984, \aap, 132, 389.

\bibitem[{Prochaska} {\it et al.}\ (2007)a]{pcw+07}
{Prochaska}, J. {\it et al.}\  2007a, In prep.

\bibitem[{Prochaska}(2006)]{pro06}
{Prochaska}, J.~X. 2006, \apj, 650, 272.

\bibitem[{Prochaska} {\it et al.}\ (2004)]{pbc+04}
{Prochaska}, J.~X. {\it et al.}\  2004, \apj, 611, 200.

\bibitem[{Prochaska} {\it et al.}\ (2006)]{pcb+07}
{Prochaska}, J.~X. {\it et al.}\  2006, ArXiv Astrophysics e-prints.

\bibitem[{Prochaska}, {Chen} \& {Bloom}(2006)]{pcb06}
{Prochaska}, J.~X., {Chen}, H.-W., and {Bloom}, J.~S. 2006, \apj, 648, 95.

\bibitem[{Prochaska} {\it et al.}\ (2005)]{gcn3971}
{Prochaska}, J.~X. {\it et al.}\  2005, GRB Coordinates Network, 3971, 1.

\bibitem[{Prochaska} {\it et al.}\ (2003)a]{pgw+03}
{Prochaska}, J.~X. {\it et al.}\  2003a, \apjl, 595, L9.

\bibitem[{Prochaska} {\it et al.}\ (2003)b]{p03_esi}
{Prochaska}, J.~X. {\it et al.}\  2003b, \apjs, 147, 227.

\bibitem[{Prochaska} {\it et al.}\ (2002)a]{pho+02}
{Prochaska}, J.~X. {\it et al.}\  2002a, \pasp, 114, 933.

\bibitem[{Prochaska} \& {Herbert-Fort}(2004)]{ph04}
{Prochaska}, J.~X. and {Herbert-Fort}, S. 2004, \pasp, 116, 622.

\bibitem[{Prochaska}, {Herbert-Fort} \& {Wolfe}(2005)]{phw05}
{Prochaska}, J.~X., {Herbert-Fort}, S., and {Wolfe}, A.~M. 2005, \apj, 635,
  123.

\bibitem[{Prochaska} {\it et al.}\ (2002)b]{pro_ion02}
{Prochaska}, J.~X. {\it et al.}\  2002b, \apj, 571, 693.

\bibitem[{Prochaska} {\it et al.}\ (2000)]{pnc+00}
{Prochaska}, J.~X. {\it et al.}\  2000, \aj, 120, 2513.

\bibitem[{Prochaska} \& {Wolfe}(1999)]{pw99}
{Prochaska}, J.~X. and {Wolfe}, A.~M. 1999, \apjs, 121, 369.

\bibitem[{Prochaska} \& {Wolfe}(2002)]{pw02}
{Prochaska}, J.~X. and {Wolfe}, A.~M. 2002, \apj, 566, 68.

\bibitem[{Prochaska} {\it et al.}\ (2007)b]{pwh+07}
{Prochaska}, J.~X. {\it et al.}\  2007b, ArXiv Astrophysics e-prints.

\bibitem[{Prochaska} {\it et al.}\ (2001)]{pro01}
{Prochaska}, J.~X. {\it et al.}\  2001, \apjs, 137, 21.

\bibitem[{Prochter} {\it et al.}\ (2006)]{ppc+06}
{Prochter}, G.~E. {\it et al.}\  2006, \apjl, 648, L93.

\bibitem[{Qian} \& {Wasserburg}(2003)]{qw03}
{Qian}, Y.-Z. and {Wasserburg}, G.~J. 2003, \apjl, 596, L9.

\bibitem[{Ramirez-Ruiz}, {Trentham} \& {Blain}(2002)]{rtb02}
{Ramirez-Ruiz}, E., {Trentham}, N., and {Blain}, A.~W. 2002, \mnras, 329, 465.

\bibitem[{Razoumov} {\it et al.}\ (2006)]{rnp+06}
{Razoumov}, A.~O. {\it et al.}\  2006, \apj, 645, 55.

\bibitem[{Reichart} \& {Price}(2002)]{rp02}
{Reichart}, D.~E. and {Price}, P.~A. 2002, ApJ, 565, 174.

\bibitem[{Richards} {\it et al.}\ (2004)]{rng+04}
{Richards}, G.~T. {\it et al.}\  2004, \apjs, 155, 257.

\bibitem[{Savaglio}(2006)]{savaglio06}
{Savaglio}, S. 2006, New Journal of Physics, 8, 195.

\bibitem[{Savaglio} \& {Fall}(2004)]{sf04}
{Savaglio}, S. and {Fall}, S.~M. 2004, \apj, 614, 293.

\bibitem[{Savaglio}, {Fall} \& {Fiore}(2003)]{sff03}
{Savaglio}, S., {Fall}, S.~M., and {Fiore}, F. 2003, ApJ, 585, 638.

\bibitem[{Sheinis} {\it et al.}\ (2000)]{sheinis00}
{Sheinis}, A.~I. {\it et al.}\  2000, in { Proc. SPIE Vol. 4008, p. 522-533,
  Optical and IR Telescope Instrumentation and Detectors, Masanori Iye; Alan F.
  Moorwood; Eds.}, 522.

\bibitem[{Shin} {\it et al.}\ (2006)]{sbp+06}
{Shin}, M.-S. {\it et al.}\  2006, ArXiv Astrophysics e-prints.

\bibitem[{Smecker-Hane} {\it et al.}\ (2002)]{scg+02}
{Smecker-Hane}, T.~A. {\it et al.}\  2002, \apj, 566, 239.

\bibitem[{Sneden}, {Gratton} \& {Crocker}(1991)]{sgc91}
{Sneden}, C., {Gratton}, R.~G., and {Crocker}, D.~A. 1991, \aap, 246, 354.

\bibitem[{Sofia} {\it et al.}\ (2006)]{sgc+06}
{Sofia}, U.~J. {\it et al.}\  2006, \apj, 636, 753.

\bibitem[{Sofia} \& {Jenkins}(1998)]{sj98}
{Sofia}, U.~J. and {Jenkins}, E.~B. 1998, \apj, 499, 951.

\bibitem[{Sollerman} {\it et al.}\ (2005)]{sof+05}
{Sollerman}, J. {\it et al.}\  2005, New Astronomy, 11, 103.

\bibitem[{Solomon} \& {Rivolo}(1989)]{sr89}
{Solomon}, P.~M. and {Rivolo}, A.~R. 1989, \apj, 339, 919.

\bibitem[{Spergel} {\it et al.}\ (2006)]{wmap06}
{Spergel}, D.~N. {\it et al.}\  2006, ArXiv Astrophysics e-prints.

\bibitem[{Stanek} {\it et al.}\ (2006)]{sgb+06}
{Stanek}, K.~Z. {\it et al.}\  2006, Acta Astronomica, 56, 333.

\bibitem[{Stanek} {\it et al.}\ (2003)]{smg+03}
{Stanek}, K.~Z. {\it et al.}\  2003, ApJ ({\it Letters}), 591, L17.

\bibitem[{Tinsley}(1979)]{tinsley79}
{Tinsley}, B.~M. 1979, \apj, 229, 1046.

\bibitem[{Totani}(1999)]{tot99}
{Totani}, T. 1999, ApJ, 511, 41.

\bibitem[{Tumlinson} {\it et al.}\ (2007)]{tpd+07}
{Tumlinson}, J. {\it et al.}\  2007, In prep.

\bibitem[{Tumlinson} {\it et al.}\ (2002)]{tsr+02}
{Tumlinson}, J. {\it et al.}\  2002, \apj, 566, 857.

\bibitem[{Vink} \& {de Koter}(2005)]{vd05}
{Vink}, J.~S. and {de Koter}, A. 2005, \aap, 442, 587.

\bibitem[{Vladilo} {\it et al.}\ (2006)]{vcl+06}
{Vladilo}, G. {\it et al.}\  2006, \aap, 454, 151.

\bibitem[{Vogt} {\it et al.}\ (1994)]{vogt94}
{Vogt}, S.~S. {\it et al.}\  1994, in { Proc. SPIE Instrumentation in Astronomy
  VIII, David L. Crawford; Eric R. Craine; Eds., Volume 2198, p. 362}, 362.

\bibitem[{Vreeswijk} {\it et al.}\ (2004)]{vel+04}
{Vreeswijk}, P.~M. {\it et al.}\  2004, \aap, 419, 927.

\bibitem[{Vreeswijk} {\it et al.}\ (2006)a]{vls+07}
{Vreeswijk}, P.~M. {\it et al.}\  2006a, ArXiv Astrophysics e-prints.

\bibitem[{Vreeswijk} {\it et al.}\ (2006)b]{vsf+06}
{Vreeswijk}, P.~M. {\it et al.}\  2006b, \aap, 447, 145.

\bibitem[{Wasserburg} \& {Qian}(2000)]{wq00}
{Wasserburg}, G.~J. and {Qian}, Y.-Z. 2000, \apjl, 538, L99.

\bibitem[{Watson} {\it et al.}\ (2006)]{wfl+06}
{Watson}, D. {\it et al.}\  2006, \apj, 652, 1011.

\bibitem[{Waxman} \& {Draine}(2000)]{wd00}
{Waxman}, E. and {Draine}, B.~T. 2000, ApJ, 537, 796.

\bibitem[{Welty} {\it et al.}\ (2006)]{welty06}
{Welty}, D.~E. {\it et al.}\  2006, \apjs, 165, 138.

\bibitem[{Welty} {\it et al.}\ (1997)]{welty97}
{Welty}, D.~E. {\it et al.}\  1997, \apj, 489, 672.

\bibitem[{Welty} {\it et al.}\ (2001)]{welty01}
{Welty}, D.~E. {\it et al.}\  2001, \apjl, 554, L75.

\bibitem[{White} {\it et al.}\ (1999)]{wnh+99}
{White}, G.~J. {\it et al.}\  1999, \aap, 342, 233.

\bibitem[{Wolf} \& {Podsiadlowski}(2006)]{wp06}
{Wolf}, C. and {Podsiadlowski}, P. 2006, ArXiv Astrophysics e-prints.

\bibitem[{Wolfe}, {Gawiser} \& {Prochaska}(2005)]{wgp05}
{Wolfe}, A.~M., {Gawiser}, E., and {Prochaska}, J.~X. 2005, \araa, 43, 861.

\bibitem[{Wolfe}, {Prochaska} \& {Gawiser}(2003)]{wpg03}
{Wolfe}, A.~M., {Prochaska}, J.~X., and {Gawiser}, E. 2003, \apj, 593, 215.

\bibitem[{Wolfe} {\it et al.}\ (1986)]{wolfe86}
{Wolfe}, A.~M. {\it et al.}\  1986, \apjs, 61, 249.

\bibitem[Woosley(1993)]{woo93}
Woosley, S.~E. 1993, ApJ, 405, 273.

\bibitem[{Woosley}(1993)]{w93}
{Woosley}, S.~E. 1993, \apj, 405, 273.

\bibitem[{Woosley} \& {Bloom}(2006)]{wb06}
{Woosley}, S.~E. and {Bloom}, J.~S. 2006, \araa, 44, 507.

\bibitem[{Woosley} \& {Heger}(2006)]{wh06}
{Woosley}, S.~E. and {Heger}, A. 2006, \apj, 637, 914.

\bibitem[{Woosley} \& {Weaver}(1995)]{ww95}
{Woosley}, S.~E. and {Weaver}, T.~A. 1995, \apjs, 101, 181.

\bibitem[{Zwaan} \& {Prochaska}(2006)]{zp06}
{Zwaan}, M.~A. and {Prochaska}, J.~X. 2006, \apj, 643, 675.

\end{thebibliography}


\clearpage

\end{document}